%% file: main.tex
\documentclass[journal]{IEEEtran}
\IEEEoverridecommandlockouts
\usepackage[utf8]{inputenc}
\usepackage{lipsum}
\input{packages}

\input{commands}

\title{MPPI-Generic: A CUDA Library for Stochastic Trajectory Optimization}

\date{\today}

\begin{document}
\author{Bogdan Vlahov$^1$, 
Jason Gibson$^1$, 
Manan Gandhi$^1$,
Evangelos A. Theodorou$^1$
\thanks{$^{1}$ Autonomous Control and Decision Systems Lab, Georgia Institute of Technology, Atlanta GA 30313 USA (\textit{Corresponding author: Bogdan Vlahov)}}}

\maketitle
\thispagestyle{plain}
\begin{abstract}
    This paper introduces a new C++/CUDA library for GPU-accelerated stochastic trajectory optimization called MPPI-Generic. It provides implementations of Model Predictive Path Integral control, Tube-Model Predictive Path Integral Control, and Robust Model Predictive Path Integral Control, and allows for these algorithms to be used across many pre-existing dynamics models and cost functions. Furthermore, researchers can create their own dynamics models or cost functions following our API definitions without needing to change the actual Model Predictive Path Integral Control code. Finally, we compare computational performance to other popular implementations of Model Predictive Path Integral Control over a variety of GPUs to show the real-time capabilities our library can allow for. Our library can be found at: \url{https://acdslab.github.io/mppi-generic-website/}
\end{abstract}

\section{Introduction}
\label{sec:intro}
As robotics and autonomy continue to grow, the choice of algorithms and methods used to plan and control complex systems in these fields start to become more selective.  
These algorithms need to be able to handle the potentially intricate dynamics found in robotic systems and allow for complex cost function representations. 
They also must be responsive to new data as environments and systems change. 
Finally, the planning and control methods used should run in real-time so that the robot can continue to perform without unnecessary pauses. 
We can see these requirements show themselves in a self-driving car example. 
The dynamics need to be aware of changing road conditions while the cost function has to capture multiple goals such as avoiding other cars, following traffic laws, and getting to the destination.
The planner has to be fast enough to react to other drivers stopping suddenly or debris on the road.

The approaches to \ac{MPC} optimization can be delineated into two types of methods: gradient-based and sampling-based.
Gradient-based methods such as \ac{iLQR} \cite{li2004iterative}, \ac{DDP} \cite{jacobson1970differential} and \ac{SQP} \cite{boggs1995sequential} generally rely on restrictions to the dynamics and cost functions such as being continuously differentiable.
But in exchange for those restrictions, they can produce controls that minimize the cost function in a computationally-efficient manner.
Sampling-based methods, such as \ac{MPPI} \cite{williams2016aggressive} or \ac{CEM} \cite{rubinstein1999cross}, can relax these requirements and allow for arbitrary functions but come at the cost of requiring many samples to properly estimate the optimal control.
These arbitrary functions remove the need for convexification or smoothing which can cause unnecessary conservatism.
One way to address this computational expense is to push the computation out of the CPU and onto a GPU, where the parallelization of sampling can be better utilized.
By taking advantage of the GPU, we make sampling-based methods usable in real-time and also provide enough samples to get optimal solutions.

First introduced in \cite{williams2016aggressive}, \ac{MPPI} is a stochastic \ac{MPC} algorithm derived using the information theoretic dualities between relative entropy and free energy. 
Experiments in \cite{williams2016aggressive} included off-road navigation using the GT-Autorally vehicle \cite{goldfain2019autorally} and demonstrated for the first time utilization of GPU for sampling-based \ac{MPC} on real hardware. 
Follow-up works included the derivation of \ac{MPPI} for the case of non-affine dynamics \cite{williams2017information,williams2018information} and extensions to multi-layer control architectures that incorporate the benefits of \ac{iLQR} to increase robustness to disturbances and model errors. 
These extensions include the Tube-based \ac{MPPI} \cite{williams2018robust}, \ac{RMPPI} \cite{williams2019model,gandhi2021robust} architectures which consists of two layers of control.
Further discussion of extensions to \ac{MPPI} is left to \cref{subsec:other_mppi_variations}.

Besides the obvious use-case of \ac{MPC}, alternative use-cases of MPPI can be found in the \ac{MBRL} literature \cite{hansen2022temporal,hansen2024tdmpc,bhardwaj2021blending,Hansen2024HierarchicalWM,qu2023rl,watson2023inferring}.
In \ac{MBRL}, \ac{MPPI} is used to seek data from a simulation environment to learn the underlying value function and corresponding policy.  Finally, while the majority of prior work on \ac{MPPI} has direct applications to planning and control for systems in robotics and autonomy such as quad-rotors \cite{pravitra2020L1,miura2024spline}, terrestrial \cite{williams2016aggressive,vlahov2024low,mohamed2022autonomous,han2023model,hannes2023efficient}, sea surface \cite{patel2023modelpredictive,liu2023data} and underwater \cite{nicolay2023enhancing} vehicles, manipulators \cite{weijia2019model}, and systems with multi-body dynamics \cite{chao2024flocking}, there are notable applications to other domains of science and engineering. These include control of HVAC systems \cite{zhiyu2023clue,ding2024multi}, chemical reactors \cite{ai2024model}, and pulse width modulation rectifiers \cite{jiayu2023finite}.



Given the far reaches of the \ac{MPPI} algorithm, it is important to provide a solid and flexible computational framework from which researchers can make use of previous advancements to push their own work forwards. That framework needs to not only provide many built-in options to allow for testing on different platforms but also allow for researchers to develop new ideas on top of. Finally, that library needs to be able to support researchers all the way to hardware deployment with the ability to run in real-time even on older hardware platforms.

With this inspiration in mind, we introduce our controls and planning optimization library, \ac{MPPI}-Generic. It is written in C++/CUDA and contains multiple dynamics and cost functions to allow for researchers to begin using them in complex robotics scenarios. In addition, it also allows for researchers to create their own dynamics or cost functions that take advantage of the GPU-accelerated controllers. It provides implementations of \ac{MPPI}\cite{williams2016aggressive}, Tube-\ac{MPPI}\cite{williams2018robust}, and \ac{RMPPI} \cite{gandhi2021robust} as well as an API for implementations of new sampling-based controller.
Initial versions of this library have already been used in a variety of hardware and software experiments \cite{vlahov2024low,patel2023modelpredictive,gandhi2021robust,gandhi2022safety,gibson2023multistep}. To the best of our knowledge, this is the first \ac{MPPI} library implementation to provide GPU-acceleration with real-time performance, multiple existing dynamics and cost functions, replaceable sampling distributions and controller algorithms, and extensibility options for researchers to create new components.

Our contributions are summarized as follows:
\begin{itemize}
    \item We provide a NVIDIA GPU-accelerated Library\footnote{Library code is located at \url{https://github.com/ACDSLab/MPPI-Generic}} for \ac{MPPI}.
    \item We provide a C++/CUDA API that allows for arbitrary dynamics and cost function definitions while maintaining high computational performance. We show simple examples of extending various parts of the library to give researchers an idea of how they could customize this library for their own needs.
    \item We show computational comparisons with other popular \ac{MPPI} libraries across a variety of computational hardware.
\end{itemize}

The rest of the paper is organized as follows: In \cref{sec:Problem_Formulation}, we provide the general problem formulation of a stochastic trajectory optimization problem, introduce various extensions to \ac{MPPI}, and discuss the building blocks of the MPPI-Generic Library.
In \cref{sec:performance}, we discuss implementation details of our library and performance parameters available to the user to tweak.
In \cref{sec:structure}, we show how to use the library through coding examples. 
In \cref{sec:benchmarks}, we show computational comparisons of our library against other implementations of \ac{MPPI}.
Finally, we state our conclusions in \cref{sec:conclusion}.

\section{Background on Stochastic Trajectory Optimization}
\label{sec:Problem_Formulation}

Consider a general nonlinear system with discrete dynamics and cost function of the following form:
\begin{align}
    \vb{x}_{t+1} &= \vb{F}\PP{\vb{x}_t, \vb{u}_t + \vb{n}_t}\label{eq:dynamics}\\
    \J(X,U) &= \phi(\vb{x}_T) + \sum_{t=0}^{T-1} \vb{\ell}\PP{\vb{x}_t, \vb{u}_t},
    \label{eq:nonlinear_control_optimization}
\end{align}
where $\vb{x} \in \R^{n_x}$ is the state, $\vb{u} \in \R^{n_u}$ is the control, $\vb{n}$ is assumed to be zero-mean Gaussian disturbances with variance $\Sigma$ in the control channel, 
$T$ is the time horizon, $X$ is a state trajectory $[\vb{x}_0, \vb{x}_1, ..., \vb{x}_T]$, $U$ is a control trajectory $[\vb{u}_0, \vb{u}_1, ..., \vb{u}_{T-1}]$, $\phi$ is the terminal cost, and $\vb{\ell}$ is the running cost. 
It is important to note that existing \ac{MPPI}-Generic dynamics assume Euler integration by default but can be modified if desired by developing a custom dynamics class.

\subsection{The MPPI Algorithm}
\label{subsec:mppi}


While \ac{MPPI} was originally derived from a path-integral approach, there have been other derivations that lead to slightly different update rules. We will briefly go over the information-theoretic derivation \cite{williams2018information} to underpin the specific update rule we use in \ac{MPPI}-Generic. We also briefly discuss other variations of \ac{MPPI} that are included in this library.

\subsubsection{Information Theoretic Derivation}
\label{subsec:info_theory_approach}

\input{info_theory_short}

\subsection{Other MPPI Modifications}
\label{subsec:other_mppi_variations}
Beyond the derivations of \ac{MPPI}, there have been other modifications that address various limiations of \ac{MPPI}.
A Tube-based \ac{MPPI} controller \cite{williams2018robust} was created in order to improve robustness to state disturbances. 
It made use of a tracking controller to track the real system back to a nominal system that ignored state disturbances resulting in large costs. 
The nominal system would use an initial state produced from the dynamics equation in \cref{eq:dynamics} for a new iteration of \ac{MPPI} instead of the real system's state estimate. 
Both the real and the nominal trajectories are calculated using \ac{MPPI} while the tracking controller was \ac{iLQR}. 
In this setup, the tracking controller would always be the one sending controls to the system and as \ac{MPPI} was not aware of the tracking controller, it could end up fighting against the tracking controller. 
In order to address that, \ac{RMPPI} was developed in \cite{williams2019model,gandhi2021robust}, which applied the tracking controller feedback within the samples \ac{MPPI} used. 
\ac{RMPPI} also chose the initial state for the nominal system using a constrained optimization problem that tried to keep the nominal state as close to the real state without causing the resulting trajectory to have a cost larger than a given cost threshold $\alpha$. 
Choosing the nominal state in this way ensured there was an upper bound on how quickly the optimal trajectory's cost could grow due to disturbance.
However, this improved robustness to state disturbances can struggle when the cost function itself changes over time due to new information.
Our library contains implementations of these algorithmic improvements to \ac{MPPI} as different controllers are the best choice in different scenarios. 

Since this first wave of publications on \ac{MPPI}, there has been a plethora of extensions and improvements. 
These include sampling efficiency improvements such as Covariance Control MPPI \cite{yin2022trajectory}, Colored MPPI \cite{vlahov2024low}, Residual MPPI \cite{wang2024residual}, Biased-MPPI \cite{trevisan2024biased}, Spline-Interpolated Model Predictive MPPI \cite{miura2024spline}, Stein Variational Guided MPPI \cite{honda2023stein},  log-\ac{MPPI} \cite{mohamed2022autonomous}, CoVO-\ac{MPC} \cite{yi2024covo}, U-\ac{MPPI} \cite{mohamed2025towards}, o-\ac{MPPI} \cite{yan2024outputsampled}, and Smooth MPPI \cite{kim2022smooth}; there have also been robustness improvements such as Tsallis-MPPI\cite{wang2021variational},  Constrained Covariance Steering MPPI \cite{balci2022constrained}, Multi-Modal MPPI \cite{zhang2024multi}, and Risk-Aware MPPI \cite{yin2023risk}.
Some of the sampling efficiency methods have also been implemented, shown in \cref{subsubsec:sampling_distributions}, but the rapid development of the community makes it nearly impossible to keep up.

\subsection{MPPI-Generic's MPPI Implementation}
In practice, \ac{MPPI}-Generic uses the update rule, \cref{eq:info_theory_opt_control_seq}, with Monte-Carlo sampling to find the optimal control sequence. 
The algorithm starts by sampling control trajectories, running each trajectory through the dynamics in \cref{eq:dynamics} to create a corresponding state trajectory, and evaluating each state and control trajectory through the cost function. 
We provide the option to calculate an importance sampling ratio calculated for the specific sampling distribution like in \cref{eq:info_theory_weighting} 
or to disable it,
\begin{align}
    \mathcal{I}\PP{V} &= \beta \ln\PP{\frac{p\PP{V}}{s\PP{V|\theta}}}
\end{align}
where $\beta \in \PCB{0,1}$. The option to disable the importance sampling weight comes from experimentation, such as on the AutoRally platform \cite{goldfain2019autorally}, where we have seen better control trajectories without it.
Finally, a weighted average of the trajectories is conducted to produce the optimal control trajectory. 
The update law for $\mathcal{U}^{*}_t$, the optimal control at time $t$, ends up looking like
\begin{align}
    \mathcal{U}^{*}_t &= \sum_{m=1}^{M} \frac{\expf{-\frac{1}{\lambda} \PP{\mathcal{J}\PP{V^m} - \lambda\mathcal{I}\PP{V^m} - \rho}}  \vb{v}^m_t}{\sum_{j=1}^{M}\expf{-\frac{1}{\lambda} \PP{\mathcal{J}\PP{V^m} - \lambda\mathcal{I}\PP{V^m} - \rho}}}
    \label{eq:mppi_update_rule}\\
    \rho &= \min_{m \in \PB{1,M}}\PP{\mathcal{J}\PP{V^m} - \lambda\mathcal{I}\PP{V^m}}
\end{align}
where $V^m$ is the $m$-th sampled control trajectory, $\vb{v}^m_t$ is the control from the $m$-th sampled trajectory at time $t$, and $\rho$ is the minimum sampled cost.
When using a Gaussian distribution, the sampled control $\vb{v}^m_t = \vb{u}_t + \epsilon^m_t$ is centered around the previous optimal control $\vb{u}_t$ and has noise $\epsilon^m_t \sim \normal{0, \Sigma}$; other distributions would have different ways to draw samples and alternative $\mathcal{I}\PP{V}$ equations.
We use $\rho$ as in \cite{williams2018information} to shift the range of the exponentiated costs to limit numerical stability issues.
Sampling in the control space ensures that the trajectories are dynamically feasible and allows us to use non-differentiable dynamics and cost functions. Pseudo-code for the algorithm is shown in \cref{alg:MPPI}.

\begin{algorithm}[ht]
\footnotesize
\LinesNumbered
\SetKwInOut{Input}{Given}
\Input{
    $\vb{F}\PP{\cdot, \cdot}$, $G\PP{\cdot, \cdot}$ $\ell\PP{\cdot, \cdot}$, $\phi\PP{\cdot}$, $\mathcal{I}\PP{\cdot}$, $M$, $I$, $T$, $\lambda$, $\Sigma$: System dynamics, system observation, running cost, terminal cost, importance sampling ratio, num. samples, num. iterations, time horizon, temperature, covariance\;
}
\SetKwInOut{Input}{Input}
\Input{
    $\vb{x}_0$, $\vb{U}$: initial state, mean control sequence\;
      }
\SetKwInOut{Input}{Output}
\Input{
    $\mathcal{U}$: optimal control sequence
}
\BlankLine
\tcp{Begin Cost sampling}
\For{$i \leftarrow 1$ \KwTo $I$}{
\For{$m \leftarrow 1$ \KwTo $M$}{
    $J^m \leftarrow 0$\;
    $\vb{x} \leftarrow \vb{x}_0$\;
    \For{$t \leftarrow 0$ \KwTo $T-1$}{
        $\vb{v}_{t} \leftarrow \vb{u_t} + \epsilon^m_t,\ \epsilon^m_t \sim \normal{0, \Sigma}$\; 
        $\vb{x} \leftarrow \vb{F}\left(\vb{x}, \vb{v}_t\right)$\;
        $\vb{y} \leftarrow G\PP{\vb{x}, \vb{v}_t}$\;
        $J^m \pe \ell\PP{\vb{y}, \vb{v}_t} - \lambda\mathcal{I}\PP{\vb{v}_t}$;
    }
    $J^m \pe \phi\left(\vb{y}\right)$
}
\tcp{Compute trajectory weights}
$\rho \leftarrow \min\left\{J^1, J^2, ..., J^M\right\}$\;
$\eta \leftarrow \sum_{m=1}^{M}\expf{-\frac{1}{\lambda}\left(J^m - \rho\right)}$\;
\For{$m \leftarrow 1$ \KwTo $M$}{
    $w^m \leftarrow \frac{1}{\eta} \expf{-\frac{1}{\lambda}\left(J^m - \rho\right)}$\;
}

\tcp{Control update}
\For{$t \leftarrow 0$ \KwTo $T-1$}{
    $\mathcal{U}_t \leftarrow \vb{u}_t + \sum_{m=1}^{M}w^m \epsilon^m_t$\; \label{alg:policy-update-line}
}
}
\caption{\ac{MPPI}}
\label{alg:MPPI}
\end{algorithm}



\subsection{Library Description}
\label{sec:library_description}
This library is made up of 6 major types of classes:
\begin{multicols}{2}
\begin{itemize}[topsep=0pt]
    \item Dynamics
    \item Cost Functions
    \item Controllers
    \item Sampling Distributions
    \item Feedback Controllers
    \item Plants
\end{itemize}
\end{multicols}
The Dynamics and Cost Function classes are self-evident and are classes describing the $\vb{F}$, $\vb{G}$, $\ell$, and $\phi$ functions from \cref{eq:dynamics,eq:observation,eq:cost}. 
The Controller class finds the optimal control sequence $\vb{U}^*$ that minimizes the cost in \cref{eq:cost} using algorithms such as \ac{MPPI}. 
The Sampling Distributions are used by the Controller class to generate the control samples used for determining the optimal control sequence. 
The Feedback Controller class determines what feedback controller if any, is used to help push the system back towards the desired trajectory computed by the Controller. 
This is required for the Tube-\ac{MPPI} and \ac{RMPPI} implementations but can even be used with the \ac{MPPI} controller as it provides feedback in between \ac{MPC} iterations.
Unless otherwise specified, the Feedback Controllers in code examples are instantiated but turned off by default. 
Finally, Plants are \iac{MPC} wrapper around a given controller and are where the interface methods in and out of the controller are generally defined. 
For example, a common-use case of \ac{MPPI} is on a robotics platform running \ac{ROS} \cite{macenski2022robot}. 
The Plant is where you would implement your \ac{ROS} subscribers to information such as state, \ac{ROS} publishers of the control output, and the necessary methods to convert from \ac{ROS} messages to \ac{MPPI}-Generic equivalents. 
Each class type has their own parameter structures which encapsulate the adjustable parameters of each specific instantiation of the class. 

\section{Performance Implementation}
\label{sec:performance}
We shall now discuss some of the performance-specific implementation details we make use of in \ac{MPPI}-Generic. 
First, we will give a brief introduction to GPU hardware and terminology followed by general GPU performance tricks useful for implementation in this library. 

\subsection{GPU Parallelization Overview}
\label{subsec:cuda_basics}
The GPU is a highly parallelizable hardware device that performs operations differently than a CPU.
The lowest level of computation in CUDA is a \emph{warp}, which consists of $32$ threads doing the same operation at every clock step \cite{nvidia2024hardware}.
These warps are grouped together to produce \emph{thread blocks}.
While the threads in a warp are computed together, the individual warps in the thread block are not guaranteed to be at the same place in the code at any given time and sometimes it can actually be more efficient to allow them to differ.
The threads in a thread block all have access a form of a local cache called \emph{shared memory}.
Like any memory shared between multiple threads on the CPU, proper mutual exclusion needs to be implemented to avoid race conditions.
Threads in a block can be given indices in 3 axes, $x$, $y$, and $z$, which we use to denote different types of parallelization within the library.
The conversion from a thread's 3D index of $(x,y,z)$ to its thread number in the block is given by \lstinline{(z * blockDim.y + y) * blockDim.x + x}.
Thread blocks can themselves be grouped into \emph{grids} and are also organized into $x$, $y$, and $z$ axes.
This is useful for large parallel operations that cannot fit within a single thread block.
The GPU code is compiled into \emph{kernels}, which can be provided arbitrary grid and block dimensions at runtime. 

It is important to briefly understand the hierarchy of memory before discussing how to improve GPU performance.
At the highest level, we have \emph{global} memory which is generally measured in \si{\giga\byte}s and very slow to access data from.
Next, we have an L2 cache in the size of \si{\mega\byte}s which can speed up access to frequently-used global data.
Then we have the L1 cache, shared memory, and CUDA texture caches.
The L1 cache and shared memory are actually the same memory on hardware and are generally several \si{\kilo\byte}s in size; they are separated by programmers explicitly using shared memory and the GPU automatically filling the L1 cache.
The CUDA texture cache is a fast read-only memory used for CUDA textures which are 2D or 3D representations of data such as a map.

\subsection{General GPU speedups}
\label{subsec:gpu_improvements_general}

When looking into writing more performant code, there are some general tricks that we  leveraged throughout our code library. The first is the use of CUDA streams \cite{harris2012how}.
By default, every call to the GPU blocks the CPU code from moving ahead.
CUDA streams allow for the asynchronous scheduling of tasks on the GPU while the CPU performs other work in the meantime. We use CUDA streams throughout in order to schedule memory transfers between the CPU and GPU as well as kernel calls and have different streams for controller optimal control and visualization computations. 

The next big tip is minimizing global memory accesses.
Global memory reading or writing can be a large bottleneck in computation time and for our library, it can be slower than the actual computations we want to do on the GPU.
The first recommendation is to move commonly-accessed data from global memory to shared memory \cite{harris2013using}.
We also use \acp{CRTP} \cite{coplien1995curiously} as our choice of polymorphism on the GPU to avoid the need of constructing and reading from a virtual method table which would be stored in global memory.

We utilize vectorized-memory \cite{luitjens2013cuda} accessing where possible.
Looking at the GPU instruction set, CUDA provides instructions to read and write in 32, 64, and 128 bit chunks, making it possible to load up to four $32$-bit floats in a single instruction.
Using these concepts, we greatly reduce the number of calls to global memory and consequently increase the speed at which our computations can run.

We also make use of hardware-defined mathematical operators called \emph{intrinsics} in some places as well to reduce computation time \cite{nvidia2024intrinsics}. 
These instrinsics are approximations of various mathematical operations such as division, $\sin$, etc. that are implemented at the hardware level, requiring many fewer clock cycles to compute. 
The trade-off is that they are approximations and can return incorrect evaluations depending on the inputs. 
As such, we limited our use of intrinsics to trigonometric functions like \lstinline{__cosf()}, \lstinline{__sinf()}, and \lstinline{__sincosf()}. We found that other intrinsics such as \lstinline{__fdividef()} or \lstinline{__expf()} when used throughout the code base cause significantly different optimal control sequence calculations. However, there are plenty of specific locations in the code where more intrinsics can be introduced without negative effects on accuracy at a future point. In addition, we try to make use of \lstinline{float}-specific methods when applicable such as \lstinline{expf()} to prevent unnecessary conversions to and from \lstinline{double}.

\begin{figure*}[htp!]
    \centering
    \includegraphics[width=\textwidth]{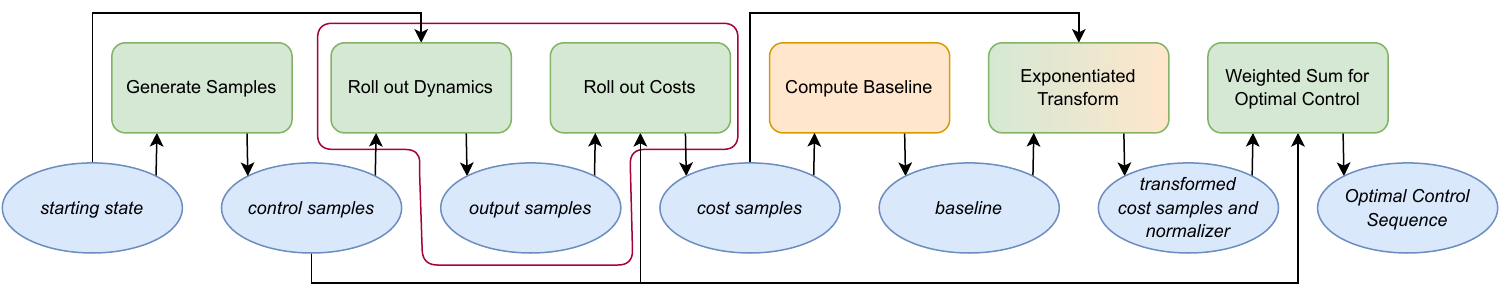}
    \caption{Diagram of the execution flow of \lstinline{computeControl()}. The blue ellipses indicate variables, the green rectangles are GPU methods, and the orange rectangles are CPU methods. The selection in purple is a single GPU kernel when using the combined kernel and separated out when using split kernels. Most of the code is run on the GPU but we found that some operations such as finding the baseline and calculating the normalizer, $\eta$, run faster on the CPU.}
    \label{fig:compute_control}
\end{figure*}

\subsection{Library-Specific Performance Optimizations}
\label{subsec:gpu_improvements_specific}
So far, we discussed optimizations that can be done for any CUDA program. However, there are further optimizations to be had in choosing how to parallelize specific components of our library.
In \cref{fig:compute_control}, we have the general steps taken every time we want to compute a new optimal control sequence in \ac{MPPI}. These same steps are also taken in Tube-\ac{MPPI} and \ac{RMPPI} though they have to be done for both the nominal and real systems.

One major performance consideration is how to parallelize the Dynamics and Cost Function calculations. 
We can either run the Dynamics and Cost Function in a combined kernel or run them in separate kernels. We describe each parallelization technique as well as the pros and cons below.
First, we introduce a slight modification to the Dynamics and Cost Functions.

\subsubsection{Intermediate Calculation Passthrough}
\label{subsec:intermediate_calcs}

When creating Cost Functions for a given Dynamics, it might be required to redo calculations already done in the Dynamics.
For example, putting a penalty on the location of the wheels of a vehicle inherently requires knowing where the wheels are located.
The wheel locations can be calculated given the position and orientation of the center of mass of the vehicle and so are not considered part of the state.
Depending on the Dynamics, the wheel locations might also be calculated as part of the state update.
To reduce unnecessary recalculations, we provide a way to pass these extra values directly from the Dynamics to the Cost Function.
We do this by slightly modifying \cref{eq:nonlinear_control_optimization} to use outputs, $\vb{y}_t$ instead of $\vb{x}_t$,
\begin{align}
    \vb{y_{t}} &= \vb{G}\PP{\vb{x}_t, \vb{u}_t} \label{eq:observation} \\
    \J(Y, U) &= \phi(\vb{y}_{T}) + \sum_{t = 0}^{T - 1}\vb{\ell}\PP{\vb{y}_t, \vb{u}_{t}} \label{eq:cost},
\end{align}
where $\vb{G}\PP{\vb{x}_t, \vb{u}_t}$ is calculated as part of the Dynamics.
For the vast majority of systems, $\vb{y}_t$ is the true state, i.e. $\vb{y}_t = \vb{x}_t$, but we have found in some cases that bringing additional calculations from the Dynamics to the Cost Function can be computationally faster than reproducing them.

\subsubsection{Split Kernel Description}
\label{subsec:split-kernel}
We start by taking the initial state and control samples and run them through the Dynamics kernel.
This kernel uses all three axes of thread parallelization for different components.
First, the $x$ dimension of the block and the grid are used to indicate which sample we are on as \lstinline{threadIdx.x + blockDim.x * blockIdx.x}.
As every sample will conduct the exact same computations, using the $x$ axis allows us to ensure that each \emph{warp} is aligned as long as the $x$ block size is chosen appropriately.
Next, the $z$ axis is used to indicate which system is being run; for \ac{MPPI}, there is only one system but Tube-\ac{MPPI} and \ac{RMPPI} use two systems, nominal and real.
As dynamics generally have different derivative computations for different states, we use the $y$ dimension, as shown in \cref{code:gpu_efficient_state_deriv}, to introduce additional parallelization within the dynamics, instead of sequential computation of each state derivative, which can lead to further performance improvements.
When our thread block's $x$ dimension is a multiple of 32, the $y$ threads are separated into different warps and \cref{code:gpu_efficient_state_deriv} improves performance.
However, when a \lstinline{switch}/\lstinline{if} statement causes threads in the same warp to follow different computations, this is known as \emph{warp divergence}, and the GPU runs the warp again to go through all code paths. 
Depending on the complexity of the branching, this can cause significant slowdowns.
\begin{code}
\begin{lstlisting}[language=C++]
int tdy = threadIdx.y;
switch(tdy) {
  case S_INDEX(X):
    xdot[tdy] = u[C_INDEX(VEL)]*cos(x[S_INDEX(YAW)]);
    break;
  case S_INDEX(Y):
    xdot[tdy] = u[C_INDEX(VEL)]*sin(x[S_INDEX(YAW)]);
    break;
  case S_INDEX(YAW):
    xdot[tdy] = u[C_INDEX(YAW_DOT)];
    break;
}
\end{lstlisting}
\caption{GPU code for the Unicycle Dynamics. This code parallelizes using the thread $y$ dimension to do each state derivative calculation in a different thread}
\label{code:gpu_efficient_state_deriv}
\end{code}
In the Dynamics kernel, we run a \lstinline{for} loop over time for each sample in which we get the current control sample, run it through the Dynamic's \lstinline{step()} method, and save out the resulting output to global memory.

Next, we look to the Cost Function ran inside its own kernel.
The reason for that is that while the Dynamics must be sequential over time, the cost function does not need to be. 
To achieve parallelization across time, we move the sample index up to the grid level and use the block's $x$ axes for time instead. 
The Cost kernel gets the control and output corresponding to the current time in its \lstinline{computeRunningCost()} or \lstinline{terminalCost()} methods, adds the cost up across time for each sample, and saves out the resulting overall cost for each sample. 
A problem that arises is the limited number of timesteps we could optimize over due to the limit of $1024$ threads in a single thread block; we address this by reusing threads to each compute multiple timesteps until we reach the desired time horizon. 
These choices brings the time to calculate the cost much closer to that of a single timestep instead of having to wait for sequential iterations of the cost if it was paired with the Dynamics kernel.

\subsubsection{Combined Kernel Description}
\label{subsec:combined-kernel}
The Combined Kernel runs the Dynamics and Cost Function methods together in a single kernel. 
This works by getting the initial state and control samples, applying the Dynamics' \lstinline{step()} to get the next state and output, and running that output through the Cost Functions' \lstinline{computeCost()} to get the cost at that time. 
This basic interaction is done in a \lstinline{for} loop over the time horizon to get the the entire state trajectory as well as the cost of the entire sample. 
We parallelize this over three axes. 
First, the $x$ and $z$ dimensions of the block and grid are used to indicate which sample and system we are on as described above in the Split Kernel's Dynamics section.
Finally, the $y$ dimension is used to parallelize within the Dynamics and Cost Functions' methods.

\subsubsection{Choosing between the Split and Combined Kernels}
\label{subsec:automatic-kernel-selection}
There are some trade-offs between the two kernel options that can affect the overall computation time. 
By combining the Dynamics and Cost Function calculations together, we keep the intermediate outputs in shared memory and do not need to save them out to global memory. 
However, we are forced to run the Cost Function sequentially in time. 
Splitting the Dynamics and Cost Function into separate kernels allows them each to use more shared memory for their internal calculations with the requirement of global memory usage to save out the sampled output trajectories. 
The Combined Kernel uses less global memory but requires more shared memory usage in a single kernel as it must contain both the Dynamics and Cost Functions' shared memory requirements.
As the number of samples grow, the number of reads and writes of outputs to global memory also grows. 
This can eventually take more time than the savings we get from running the Cost Function in parallel across time, even when using vectorized-memory reads and writes. 

In order to address these trade-offs, we implemented both kernel approaches in our library and automatically choose the most appropriate kernel in the Controller constructor using \lstinline{chooseAppropriateKernel()}. The automatic kernel selection is done by running both the combined and split kernels multiple times, measuring the computation time of each, and choosing the fastest option. As the combined kernel potentially uses more shared memory than the split kernel, we also check to see if the amount of shared memory used is below the GPU's shared memory hardware limit; if it is not, we default to the split kernel approach. We also allow the user to overwrite the automatic kernel selection through the use of the \lstinline{setKernelChoice()} method.

\subsubsection{Weight Transform and Update Rule Kernels}
\label{subsec:exponentiation-and-weighted-update}
Once the costs of each sample trajectory is calculated, we bring these costs back to the CPU in order to find the baseline, $\rho$. 
The baseline is calculated by finding the minimum cost of all the sample trajectories; it is subtracted out during the exponentiation stage as it has empirically led to better optimization performance.
When the number of samples is only in the thousands, we found that the copy to the CPU to do the baseline search is faster than doing so on the GPU. 
The costs are then exponentiated on the GPU and the normalizer, $\eta$, is calculated back on the CPU before doing the final optimal control calculation.


\subsection{Performance Recommendations}
\label{subsec:performance_recs}

\begin{code*}[ht!]
\begin{lstlisting}[language=C++,escapechar=~]
#include <mppi/controllers/MPPI/mppi_controller.cuh> ~\label{line:headers}~
#include <mppi/cost_functions/cartpole/cartpole_quadratic_cost.cuh>
#include <mppi/dynamics/cartpole/cartpole_dynamics.cuh>
#include <mppi/feedback_controllers/DDP/ddp.cuh>

const int NUM_TIMESTEPS = 100; ~\label{line:num_timesteps}~
const int NUM_ROLLOUTS = 2048; ~\label{line:num_samples}~
using DYN_T = CartpoleDynamics; ~\label{line:start_abbreviations}~
using COST_T = CartpoleQuadraticCost; ~\label{line:mid_abbreviations_1}~
using FB_T = DDPFeedback<DYN_T, NUM_TIMESTEPS>; ~\label{line:mid_abbreviations_2}~
using SAMPLING_T = mppi::sampling_distributions::GaussianDistribution<DYN_T::DYN_PARAMS_T>; ~\label{line:mid_abbreviations_3}~
using CONTROLLER_T = VanillaMPPIController<DYN_T, COST_T, FB_T, NUM_TIMESTEPS, NUM_ROLLOUTS, SAMPLING_T>; ~\label{line:mid_abbreviations_4}~
using CONTROLLER_PARAMS_T = CONTROLLER_T::TEMPLATED_PARAMS; ~\label{line:end_abbreviations}~

int main(int argc, char** argv) {
  float dt = 0.02; ~\label{line:set_dt}~
  std::shared_ptr<DYN_T> dynamics = std::make_shared<DYN_T>();  // set up dynamics ~\label{line:dynamics_variable}~
  std::shared_ptr<COST_T> cost = std::make_shared<COST_T>();    // set up cost ~\label{line:cost_variable}~
  // set up feedback controller
  std::shared_ptr<FB_T> fb_controller = std::make_shared<FB_T>(dynamics.get(), dt); ~\label{line:feedback_variable}~
  // set up sampling distribution
  SAMPLING_T::SAMPLING_PARAMS_T sampler_params; ~\label{line:sampler_params_1}~
  std::fill(sampler_params.std_dev, sampler_params.std_dev + DYN_T::CONTROL_DIM, 1.0); ~\label{line:sampler_params_2}~
  std::shared_ptr<SAMPLING_T> sampler = std::make_shared<SAMPLING_T>(sampler_params); ~\label{line:sampling_variable}~
  // set up MPPI Controller
  CONTROLLER_PARAMS_T controller_params; ~\label{line:controller_params_1}~
  controller_params.dt_ = dt; ~\label{line:controller_params_2}~
  controller_params.lambda_ = 1.0; ~\label{line:controller_params_3}~
  controller_params.dynamics_rollout_dim_ = dim3(64, DYN_T::STATE_DIM, 1); ~\label{line:controller_params_4}~
  controller_params.cost_rollout_dim_ = dim3(NUM_TIMESTEPS, 1, 1); ~\label{line:controller_params_5}~
  std::shared_ptr<CONTROLLER_T> controller = std::make_shared<CONTROLLER_T>(
      dynamics.get(), cost.get(), fb_controller.get(), sampler.get(), controller_params); ~\label{line:controller_variable}~
  
  DYN_T::state_array x = dynamics->getZeroState();  // set up initial state ~\label{line:init_state}~
  controller->computeControl(x, 1);                 // calculate control ~\label{line:optimize_control}~
  auto control_sequence = controller->getControlSeq();
  std::cout << "Control Sequence:\n" << control_sequence << std::endl;
  return 0;
}
\end{lstlisting}
\caption{Minimal Example to print out optimal control sequence for a cartpole system.}
\label{code:min_example_no_mpc}
\end{code*}

The performance capabilities of the \ac{MPPI}-Generic library is highly dependent on the choices of block sizes for the various kernels, Dynamics, and Cost Functions. There is no one choice to be made so instead, we provide some general rules of thumbs that we have seen work more often than not.
\begin{itemize}
    \item Set the thread block $x$ dimension size of both the Dynamics and Cost kernels to be a multiple of 32. This allows all the threads in a warp to do the same operation. We have seen in some cases that lowering the thread block $x$ dimension to 16 can provide some improvements but this has not been true for the majority of dynamics and cost functions.
    \item Set the thread block $y$ dimension for Dynamics equal to the state dimension and Cost Functions to 1. This depends on whether the Dynamics/Cost Function are utilizing the multi-threading capability available to them but for most of the pre-defined Dynamics, they are using the $y$ axis of the thread block to do dynamics updates. Meanwhile, most Cost Functions are not taking advantage of the parallelization as they return a single value, the cost.
    \item Keep Dynamics and Cost Function kernels' thread block sizes low. The limit to the size of a thread block is currently $1024$ so it might be tempting to fill that. However, there are various synchronization points in these kernels to ensure that data has been loaded before being used. As the thread block size increases, more time is spent waiting at these synchronization points in the Dynamics and Cost Function kernels.
    \item When developing your own Dynamics, it can be pertinent to make the output dimensions divisible by 2 or 4. This can done by adding extraneous values to the \lstinline{enum}. This allows the code to use more efficient memory-loading GPU instructions discussed in \cref{subsec:gpu_improvements_general} and reduce the number of memory calls by a factor of at least 2. One might be tempted to also ensure that the control dimension is divisible by 4. However, the control dimension is also used to generate samples so while there may be improvements to the efficiency of reading memory, it will also slow down generating samples.
    \item Minimize the occurrence of warp divergence in your GPU code when possible. The most common way these occur is by using \lstinline{if} statements based on values that might be different for individual threads in a warp. In some cases, converting the desired \lstinline{bool} expression into a \lstinline{float} will be faster while still providing the desired branching. 
\end{itemize}

\section{API Structure}
\label{sec:structure}
We describe the library API and its usage in three stages. 
At the beginner level, the desire is to use a provided Dynamics, Cost Function, and Controller to control a system.
This requires the least amount of code writing on the part of the user as most of the code is already provided.
The library user would only need to write a specific Plant class to properly interface with whatever system they are wanting to control and the executable which sets up and runs the controller itself.
The intermediate level is where the user might want to implement a dynamics model or cost function that does not exist in the base library.
Finally, at the advanced level, we show how to implement a custom Feedback Controller, Controller, or Sampling Distribution.

\subsection{Beginner}
\label{subsec:beginner}

\begin{code*}[ht!]
\begin{lstlisting}[language=C++,escapechar=~]
#pragma once
#include <mppi/core/base_plant.hpp>
#include <mppi/dynamics/cartpole/cartpole.cuh>

template <class CONTROLLER_T> class SimpleCartpolePlant : public class BasePlant<CONTROLLER_T> {
public:
  using control_array = typename CartpoleDynamics::control_array;
  using state_array = typename CartpoleDynamics::state_array;
  using output_array = typename CartpoleDynamics::output_array;

  SimpleCartpolePlant(std::shared_ptr<CONTROLLER_T> controller, int hz, int optimization_stride)~\label{line:plant_constructor}~
  : BasePlant<CONTROLLER_T>(controller, hz, optimization_stride) {
    system_dynamics_ = std::make_shared<CartpoleDynamics>();
  }

  void pubControl(const control_array& u) {~\label{line:pub_control}~
    state_array state_derivative;
    output_array dynamics_output;
    state_array prev_state = current_state_;
    float t = this->state_time_;
    float dt = this->controller_->getDt();
    system_dynamics_->step(prev_state, current_state_, state_derivative, u, dynamics_output, t, dt);
    current_time_ += dt;~\label{line:plant_update_time}~
  }

  void pubNominalState(const state_array& s) {}
  void pubFreeEnergyStatistics(MPPIFreeEnergyStatistics& fe_stats) {}
  int checkStatus() { return 0; }
  double getCurrentTime() { return current_time_; }
  double getPoseTime() { return this->state_time_; }
  
  state_array current_state_;
protected:
  std::shared_ptr<CartpoleDynamics> system_dynamics_;
  double current_time_ = 0.0;
}
\end{lstlisting}
\caption{Basic Plant implementation that interacts with a virtual Cartpole dynamics system stored within the Plant.}
\label{code:cartpole_plant}
\end{code*}

We start by showing an example of just using a single iteration of \ac{MPPI} to produce an optimal control sequence in \cref{code:min_example_no_mpc}. 
At the beginning of the example (\cref{line:start_abbreviations,line:mid_abbreviations_1,line:mid_abbreviations_2,line:mid_abbreviations_3,line:mid_abbreviations_4,line:end_abbreviations}), we create aliases such \lstinline{DYN_T} for \lstinline{CartpoleDynamics} to keep the code fairly succinct. 
The Feedback Controller is not used for this example but it is required to exist. 
The Sampling Distribution and Controller have parameters that need to be set; their use spans from algorithmic to performance -- \cref{line:sampler_params_1,line:sampler_params_2,line:controller_params_1,line:controller_params_2,line:controller_params_3,line:controller_params_4,line:controller_params_5} respectively. The \lstinline{std_dev}, \lstinline{dt_}, and \lstinline{lambda_} are parameters affecting the \ac{MPPI} update rule whereas the \lstinline{dynamics_rollout_dim_} and \lstinline{cost_rollout_dim_} parameters adjust how fast the \ac{MPPI} algorithm is computed using recommendations from \cref{subsec:performance_recs}. 
Once the components are initialized, we create an instance of the \ac{MPPI} Controller, passing the Dynamics, Cost Function, Feedback Controller, Sampling Distribution, and controller parameters to the constructor. 
On \cref{line:init_state}, we create an initial zero state for the dynamics  using \lstinline{getZeroState()} and compute an optimal control sequence with \lstinline{computeControl()} on \cref{line:optimize_control}. 
We return the control sequence as an \lstinline{Eigen::Matrixf} with $n_u$ rows and $T$ columns using \lstinline{getControlSeq()} to print out. 

When using \ac{MPPI} in \iac{MPC} fashion, we need to use a Plant wrapper around our controller.
The Plant houses methods to obtain new data such as state, calculate the optimal control sequence at a given rate using the latest information available, and provide the latest control to the external or ground truth system while providing the necessary tooling to ensure there are no race conditions. 
As this class provides the interaction between the algorithm and the actual system, it is a component that has to be modified for every use case.
For \cref{code:cartpole_plant}, we implement a plant (\lstinline{SimpleCartpolePlant}) inheriting from \lstinline{BasePlant} that contains the ground truth system completely internal to the class.
Specifically, our plant runs the external dynamics inside \lstinline{pubControl()} in order to produce a new state. We then call \lstinline{updateState()} at a different fixed rate from the controller re-planning rate to show that the capability of the code base.
\lstinline{SimpleCartpolePlant} instantiates a \lstinline{CartpoleDynamics} object in its constructor, overwrites the required virtual methods from \lstinline{BasePlant}, and sets up the dynamics update to occur within \lstinline{pubControl()}. 
Looking at the constructor on \cref{line:plant_constructor}, we pass a shared pointer to a Controller, an integer representing the controller replanning rate, and the minimum optimization stride, before creating our stand-in system dynamics.
In a new \ac{MPC} iteration, we shift the start of the mean control sequence we sample around to the maximum between the number of timesteps since the last optimization and the minimum optimization stride.
For use in \ac{MPC}, we recommend setting the minimum optimization stride to $1$.
\lstinline{pubControl()} on \cref{line:pub_control} is where we send the control to the system.
In this case, we create necessary extra variables to pass the current state $\vb{x}_t$ and control $\vb{u}_t$ as \lstinline{prev_state} and \lstinline{u} respectively to the Dynamics' \lstinline{step()} method to get the next state, $\vb{x}_{t+1}$, in the variable \lstinline{current_state_}. 
We also update the current time on \cref{line:plant_update_time} to show the system has moved forward in time. 
Looking at this class, an issue arises as the Controller it is templated upon might not use \lstinline{CartpoleDynamics} as its Dynamics class. 
This is easily remedied by replacing any reference to \lstinline{CarpoleDynamics} with \lstinline{CONTROLLER_T::TEMPLATED_DYNAMICS} to make this Plant work with the Dynamics used by the instantiated Controller.

Now that we have written our specialized Plant class, we can make some modifications to \cref{code:min_example_no_mpc} to use the controller in \iac{MPC} fashion. 
For this example, we would run a simple \lstinline{for} loop that calls the Plant's \lstinline{runControlIteration()} and \lstinline{updateState()} methods to simulate a receiving a new state from the system and then calculating a new optimal control sequence from it, replacing \cref{line:optimize_control}. 
The \lstinline{updateState()} method calls \lstinline{pubControl()} internally so the system state and the current time would update at each \lstinline{for} loop iteration. 
For real-time scenarios, the \lstinline{runControlLoop()} Plant method can be launched in a separate thread and calls \lstinline{runControlIteration()} internally at the specified re-planning rate.

\subsection{Intermediate}
\label{subsec:intermediate}
In the previous section, much of the underlying structure of the library was glossed over. 
As we get to implementing our own Dynamics or Cost Functions however, there are some basic principles to go over. 
This library is running code on two different devices, the CPU and the GPU. 
Some classes, such as Plants and Controllers, do not have methods that need to run on both whereas other classes like the Dynamics and Cost Functions do. 
The GPU can do many computations in parallel but in order to properly utilize it, that can require different code than what runs on the CPU. 
As such, when looking at implementing a new Dynamics or Cost Function, there are some methods that have to be implemented as two very similar functions, once for the CPU and once for the GPU.

\subsubsection{Custom Dynamics}
\label{intermediate:dynammics}
The first thing that needs to be implemented for a new Dynamics class is a new parameter structure. 
We implemented a dictionary-like structure in the form of \lstinline{enum} to define the state, control, and output vectors and these dictionaries are stored in the parameter. For example, in \cref{code:dyn_params}, we show a basic parameter structure implemented for a unicycle model.

\begin{code}[ht!]
\begin{lstlisting}[language=C++,firstline=6,lastline=27,escapechar=~]
struct UnicycleParams : public DynamicsParams {
  enum class StateIndex : int {
    X = 0,
    Y,
    YAW,
    NUM_STATES
  };

  enum class ControlIndex : int {
    VEL = 0,
    YAW_DOT,
    NUM_CONTROLS
  };

  enum class OutputIndex : int {
    X = 0,
    Y,
    YAW,
    NUM_OUTPUTS
  };
};
\end{lstlisting}
\caption{Simple parameter structure implementation for a unicycle model}
\label{code:dyn_params}
\end{code}

By implementing these \lstinline{enum}, we can use these state names later on in the Dynamics and Cost Function to make it clear what state we are referring to. The ending values of \lstinline{NUM_STATES}, \lstinline{NUM_CONTROLS}, and \lstinline{NUM_OUTPUTS} are also used to determine the size of each of the resulting dimensions for creating statically-sized \lstinline{Eigen::Matrixf} data types such as \lstinline{state_array} and \lstinline{control_array}.

\begin{code}[ht!]
\begin{lstlisting}[language=C++,firstline=28,lastline=49,escapechar=~]
class Unicycle : public Dynamics<Unicycle, UnicycleParams> {
public:
  using PARENT_CLASS = Dynamics<Unicycle, UnicycleParams>;

  std::string getDynamicsModelName() const override {
    return "Unicycle";
  }

  Unicycle(cudaStream_t stream = nullptr) 
  : PARENT_CLASS(stream) {}

  void computeStateDeriv(~\label{line:cpu_compute_state_deriv}~
      const Eigen::Ref<const state_array>& x,
      const Eigen::Ref<const control_array>& u,
      Eigen::Ref<state_array> x_dot);

  __device__ inline void computeStateDeriv(~\label{line:gpu_compute_state_deriv}~
      float* x, float* u,
      float* x_dot, float* theta_s);

  state_array stateFromMap(const std::map<std::string, float>& map) override;
};
\end{lstlisting}
\caption{All the methods that need to overwritten in a custom Dynamics class}
\label{code:dyn_overritten_methods}
\end{code}

Next, we implement the necessary overwritable methods. 
These methods are shown in \cref{code:dyn_overritten_methods} and start with \lstinline{getDynamcisModelName()} which returns the name of the Dynamics model. 
Next are the CPU and GPU versions of \lstinline{computeStateDeriv()} on \cref{line:cpu_compute_state_deriv,line:gpu_compute_state_deriv} respectively. 
The CPU and GPU versions are differentiated firstly by the \lstinline{__device__} keyword at the front of the GPU code to designate that method only runs on the GPU.
Next, we also only use \lstinline{Eigen::Matrixf} data-types on the CPU and raw float pointers on the GPU. 
The \lstinline{computeStateDeriv()} both would implement the following dynamics,
\begin{align}
    \dot{x} &= u_0 * cos\PP{\psi} \\
    \dot{y} &= u_0 * sin\PP{\psi} \\
    \dot{\psi} &= u_1,
\end{align}
 as shown in \cref{code:dyn_stateDeriv}. 
 Note the use of the \lstinline{enum} we created earlier with the \lstinline{S_INDEX()} and \lstinline{C_INDEX()} macros.
 This allows us to not need to know the order of the states or controls in the underlying data type and more importantly, these same \lstinline{enum} can also be used in Cost classes to ensure that compatibility with multiple dynamics as long as the dynamics define the necessary \lstinline{enum} values.
 
\begin{code}[ht!]
\begin{lstlisting}[language=C++,firstline=5,lastline=7,escapechar=~]
xdot[S_INDEX(X)] = u[C_INDEX(VEL)] * cos(x[S_INDEX(YAW)]);
xdot[S_INDEX(Y)] = u[C_INDEX(VEL)] * sin(x[S_INDEX(YAW)]);
xdot[S_INDEX(YAW)]   = u[C_INDEX(YAW_DOT)];
\end{lstlisting}
\caption{\lstinline{computeStateDeriv()} implementation for a Unicycle dynamics}
\label{code:dyn_stateDeriv}
\end{code}

An important note is that there are separate instances of the class allocated on the CPU and GPU for dynamics and cost functions.
In order to update the GPU version, you must overwrite \lstinline{paramsToDevice()} to copy from the CPU side class to the GPU side class.
Parameter structures have a default implementation for this, but if helper classes are added, their corresponding copy methods will need to be called.

Finally, the last method to be overwritten is \lstinline{stateFromMap()}
This method is used to translate \lstinline{std::map} of state names and values into the Dynamics' corresponding state vector. This method ends up being useful for the Plant, especially when different parts of the state can come in at different rates.

The methods listed above are just the beginning of the Dynamics API customization to get started. 
More advanced customization includes changing the default choice of integration scheme from Euler integration to more accurate integration schemes such as Runge-Kutta or implicit integration if needed, adjusting what is considered the zero state and control for the Dynamics, and adjusting how to linearly interpolate between states.

\subsubsection{Custom Cost Function}
\label{subsec:custom_cost}
\begin{code}[h!]
\begin{lstlisting}[language=C++,firstline=6,lastline=33,escapechar=~]
struct UnicycleCostParams : public CostParams<Unicycle::CONTROL_DIM>~\label{line:cost_params}~
{ float width = 1.0; float coeff = 10.0; };

class UnicycleCost : public Cost<UnicycleCost, UnicycleCostParams, Unicycle::DYN_PARAMS_T>
{
public:
  using PARENT_CLASS = Cost<UnicycleCost, UnicycleCostParams, Unicycle::DYN_PARAMS_T>;
  using DYN_P = PARENT_CLASS::TEMPLATED_DYN_PARAMS;~\label{line:cost_alias_for_dyn_params}~

  UnicycleCost(cudaStream_t stream = nullptr);

  std::string getCostFunctionName() 
  { return "Unicycle Cost"; }

  float computeStateCost(
    const Eigen::Ref<const output_array> y, 
    int t, int* crash_status);

  float terminalCost(
    const Eigen::Ref<const output_array> y);

  __device__ float computeStateCost(float* y, 
    int t, float* theta_c, int* crash_status);

  __device__ float terminalCost(float* y, 
    float* theta_c);
};
\end{lstlisting}
\caption{Basic Cost Function Parameter Structure and Class Implementation for the Unicycle Dynamics}
\label{code:unicycle_cost_class}
\end{code}

If we make a Dynamics class, we also generally need to make a corresponding Cost. 
While a default quadratic cost implementation that allows for use with any dynamics exists, we implement a quick Cost class to show what methods to overwrite. 
For this example, we choose to think of our unicycle on a road of some width pointing in the $x$ direction. 
We want to keep the unicycle on the road so we penalize the absolute $y$ position linearly up to the road width and penalize it quadratically if it goes outside the road. 

To do this, we first construct a parameter structure for the cost class. 
It inherits from the base CostParams class and needs to know the \lstinline{CONTROL_DIM} and have a variable for the width of the road and coefficient for the cost. 
From there, we implement a basic Cost class that has to overwrite \lstinline{computeStateCost()} and \lstinline{terminalCost()} on both the CPU and GPU sides. 
It also creates a default constructor and name method \lstinline{getCostFunctionName()}. 
This basic implementation is shown in \cref{code:unicycle_cost_class}. 
Notice that we created an alias for our dynamics parameter's \lstinline{struct} on \cref{line:cost_alias_for_dyn_params}. 
This is to allow us to use the \lstinline{enum} defined there inside our cost function as well. 
Other things to note are that again, our cost is based on the output rather than the state in \lstinline{computeStateCost()}. 
For our simple Unicycle Dynamics, the output is the same as the state.

\begin{code}[hb!]
\begin{lstlisting}[language=C++,firstline=9,lastline=16,escapechar=~]
  float cost = 0;
  float y_abs = abs(y[O_IND_CLASS(DYN_P, Y)]);
  if (y_abs < this->params_.width) {
    cost = y_abs;
  } else { // Quadratic cost outside the road width
    cost = y_abs * y_abs;
  }
  return this->params_.coeff * cost;
\end{lstlisting}
\caption{Basic State Cost Implementation for the Unicycle Dynamics}
\label{code:unicycle_state_cost}
\end{code}

In \cref{code:unicycle_state_cost}, the cost function described previously is implemented. We can use \lstinline{enum} macros such as \lstinline{O_IND_CLASS()} when we are outside of the Dynamics class to still find the appropriate output value. This code can be implemented in both the CPU and GPU \lstinline{computeStateCost()} methods and we choose \lstinline{terminalCost()} to return $0$.

These new Dynamics and Cost are easily incorporated into our previous controller examples and that should be enough to get most people started on using this library. There are more options in the Cost and Dynamics classes to improve GPU performance but those are left to \cref{sec:performance}.

\subsection{Advanced}
\label{subsec:advanced}
In this final subsection, we show off how to customize Controllers, Feedback Controllers, and Sampling Distributions. Customizing these allows users to create new sampling-based controllers or change the sampling distributions used in specific scenarios.

\subsubsection{Feedback Controllers}
\label{subsubsec:feedback}
\begin{figure}
    \centering
    \includegraphics[width=0.8\linewidth]{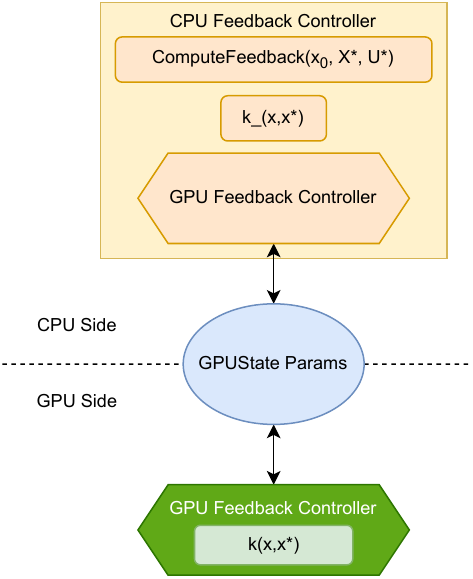}
    \caption{Feedback Controller API Diagram. We have a GPUState-based parameter structure that contains things like the feedback gains of \ac{iLQR}. The GPU Feedback Controller class exists both on the GPU (shown in green) and CPU (shown in orange) and contains the GPUState as well as a method $k(x,x^*)$ to calculate the feedback control on the GPU. The CPU Feedback Controller class (shown in yellow) is a wrapper around the GPU Feedback class that has a CPU method to calculate $k(x, x^*)$ as well as a method, \lstinline{computeFeedback()}, to recompute the feedback gains.}
    \label{fig:feedback_diagram}
\end{figure}
Feedback Controllers are useful even when using \ac{MPPI} and become necessary for Tube-\ac{MPPI} and \ac{RMPPI}. The only Feedback Controller currently implemented is \ac{iLQR} but we show how to construct a simple PID controller for each state/control combination.
\begin{code}[ht!]
\begin{lstlisting}[language=C++,firstline=7,lastline=16,escapechar=~]
template <class DYN_T>
class PIDState : public GPUState
{
  using STATE_DIM = DYN_T::STATE_DIM;
  using CONTROL_DIM = DYN_T::CONTROL_DIM;
  float p[STATE_DIM * CONTROL_DIM] = {0.0};
  float i[STATE_DIM * CONTROL_DIM] = {0.0};
  float d[STATE_DIM * CONTROL_DIM] = {0.0};
  float dt;
};
\end{lstlisting}
\caption{PID Controller Parameter Structure containing the $p$, $i$, and $d$ feedback matrices as well as the $\Delta t$ used for integral and derivative calculations.}
\label{code:pid_params}
\end{code}

Before the example Feedback Controller, we need to go over the code structure for Feedback Controllers as this is very different from the previous examples of Dynamics and Cost Functions.
This is due to the fact that the feedback controller needs to be usable on the GPU but might have different memory requirements on the CPU side.
Using \ac{iLQR} as an example in \cref{fig:feedback_diagram}, calculating the feedback on the GPU would just require the feedback gains to be sent but calculating the feedback gains requires access to the Dynamics and their Jacobians as well as a Cost Function and its Jacobian.
To handle this discrepancy in computational workloads, each Feedback Controller is made up of two parts: a GPU feedback class which contains the bare necessities to calculate the feedback, and a CPU wrapper which can do things like recomputing the gains to a new desired trajectory.

Now let us show how to utilize the Feedback Controller API to create a new PID controller. We start by constructing the parameter structure for the GPU Feedback class, shown in \cref{code:pid_params}. Note that it is templated off of the Dynamics so that it can create the appropriately-sized arrays for each gain.
Since this is a generic PID controller, we use the full feedback matrix representation for each type of gain. 
Next, we need to implement the GPU Feedback Controller for the PID in \cref{code:pid_gpu}. 
The only method that is required to be overwritten is the \lstinline{k()} feedback method on \cref{line:pid_gpu_k} but as PIDs require some history to calculate the $i$ and $d$ portions, we request extra memory by setting the \lstinline{SHARED_MEM_REQUEST_BLK_BYTES} variable on \cref{line:pid_shared_mem}.
This variable allows for extra shared memory per trajectory sample which is necessary for keeping track of history on the GPU.
\begin{code}[ht!]
\begin{lstlisting}[language=C++,firstline=18,lastline=48,escapechar=~]
template <class DYN_T>
class gpuPID : public GPUFeedbackController<gpuPID<DYN_T>, DYN_T, PIDState<DYN_T>> {
public:
  static const int SHARED_MEM_REQUEST_BLK_BYTES = DYN_T::STATE_DIM * 2;~\label{line:pid_shared_mem}~

  __device__ void k(const float* x_act,~\label{line:pid_gpu_k}~
                    const float* x_goal, const int t, 
                    float* theta, float* u_fb) {
  int tid = threadIdx.x + blockDim.x * threadIdx.z;
  float* e_int = theta[(tid*2+0) * DYN_T::STATE_DIM];
  float* pre_e = theta[(tid*2+1) * DYN_T::STATE_DIM];
  float e_der[DYN_T::STATE_DIM];
  float curr_e[DYN_T::STATE_DIM];
  for (int i = 0; i < DYN_T::STATE_DIM; i++) {
    curr_e[i] = (x_act[i] - x_goal[i]);
    e_der[i] = (curr_e[i] - pre_e[i]) / this->state_.dt;
    e_int[i] += curr_e[i] * this->state_.dt;
    pre_e[i] = curr_e[i];
  }

  // start calculating control
  for (int i = 0; i < DYN_T::CONTROL_DIM; i++) {
    for (int j = 0; j < DYN_T::STATE_DIM; j++) {
      int fb_idx = i + j * DYN_T::CONTROL_DIM;
      u_fb[i] = this->state_.p[fb_idx] * curr_e[j];
      u_fb[i] += this->state_.i[fb_idx] * e_int[j];
      u_fb[i] += this->state_.d[fb_idx] * e_der[j];
    }
  }
}
};
\end{lstlisting}
\caption{PID GPU Implementation showing how to implement $k(x,x^*)$ and request shared memory for the integral and derivative of states.}
\label{code:pid_gpu}
\end{code}

\begin{code*}[ht!]
\begin{lstlisting}[language=C++,firstline=51,lastline=92,escapechar=~]
template <class DYN_T, int NUM_TIMESTEPS>
class PIDFeedback : public FeedbackController<gpuPID<DYN_T>, PIDParams, NUM_TIMESTEPS> {
public:
    using PARENT_CLASS = FeedbackController<gpuPID<DYN_T>, PIDParams, NUM_TIMESTEPS>;
    using control_array = typename PARENT_CLASS::control_array;
    using state_array = typename PARENT_CLASS::state_array;
    using state_trajectory = typename PARENT_CLASS::state_trajectory;
    using control_trajectory = typename PARENT_CLASS::control_trajectory;
    using INTERNAL_STATE_T = typename PARENT_CLASS::TEMPLATED_FEEDBACK_STATE;
    using feedback_gain_matrix = typename DYN_T::feedback_matrix;

    PIDFeedback(float dt = 0.01, int num_timesteps = NUM_TIMESTEPS, cudaStream_t stream = 0) 
          : PARENT_CLASS(dt, num_timesteps, stream) {
        this->gpu_controller_->getFeedbackStatePointer()->dt = dt;
    }

    control_array k_(const Eigen::Ref<const state_array>& x_act, const Eigen::Ref<const state_array>& x_goal,~\label{line:pid_cpu_k}~
                     int t, INTERNAL_STATE_T& fb_state) {
        Eigen::Map<feedback_gain_matrix> P_gain(&fb_state.p);
        Eigen::Map<feedback_gain_matrix> I_gain(&fb_state.i);
        Eigen::Map<feedback_gain_matrix> D_gain(&fb_state.d);
        state_array curr_e = x_act - x_goal;
        state_array der_e = (curr_e - prev_e) / this->dt_;
        e_int_ += (curr_e) * this->dt_;
        prev_e_ = curr_e;
        control_array output = P_gain * curr_e + I_gain * e_int_ + D_gain * der_e;
        return output;
    }

    void computeFeedback(const Eigen::Ref<const state_array>& init_state,
                         const Eigen::Ref<const state_trajectory>& goal_traj,
                         const Eigen::Ref<const control_trajectory>& control_traj)  {}

protected:
    state_array e_int_ = state_array::Zero();
    state_array prev_e_ = state_array::Zero();
};
\end{lstlisting}
\caption{PID CPU Implementation. It shows the necessary type aliases and how to compute the feedback control in \lstinline{k_()}.}
\label{code:pid_cpu}
\end{code*}

After writing the GPU feedback class, we now just have to write the CPU feedback class, shown in \cref{code:pid_cpu}. We have two methods to overwrite here in \lstinline{computeFeedback()} and \lstinline{k_()}. \lstinline{computeFeedback()} is used to calculate new feedback gains given a new trajectory. For our simple PID class, we stick to constant PID gains so this can be left empty. The \lstinline{k_()} method on \cref{line:pid_cpu_k} is just the CPU version of the feedback calculation. For the CPU version of the PID, we can now use member variables to hold the history required to calculate the $i$ and $d$ portions.

\subsubsection{Controller}
\label{subsubsec:controller}
\begin{code}[hb!]
\begin{lstlisting}[language=C++,firstline=13,lastline=16,escapechar=~]
template <int STATE_DIM, int CONTROL_DIM, int MAX_T>
struct CEMParams : public ControllerParams<STATE_DIM, CONTROL_DIM, MAX_T> {
  float top_k_percentage = 0.10f;
};
\end{lstlisting}
\caption{\acs{CEM} Controller Parameter Structure containing the percentage of elite samples.}
\label{code:cem_controller_params}
\end{code}

\begin{code*}[ht!]
\begin{lstlisting}[language=C++,firstline=48,lastline=90,escapechar=~]
void computeControl(const Eigen::Ref<const state_array>& state, int optimization_stride = 1) override {
  // Send the initial condition to the device
  HANDLE_ERROR(cudaMemcpyAsync(this->initial_state_d_, state.data(), DYN_T::STATE_DIM * sizeof(float),
                               cudaMemcpyHostToDevice, this->stream_));
  for (int opt_iter = 0; opt_iter < this->getNumIters(); opt_iter++) {
    this->copyNominalControlToDevice(false); // Send the optimal control sequence to the device
    this->sampler_->generateSamples(optimization_stride, opt_iter, this->gen_, false); // Generate noise data
    
    // Calculate state trajectories and costs from sampled control trajectories
    mppi::kernels::launchSplitRolloutKernel<DYN_T, COST_T, SAMPLING_T>(
        this->model_->model_d_, this->cost_->cost_d_, this->sampler_->sampling_d_, this->getDt(),
        this->getNumTimesteps(), NUM_SAMPLES, this->getLambda(), this->getAlpha(), this->initial_state_d_,
        this->output_d_, this->trajectory_costs_d_, this->params_.dynamics_rollout_dim_,
        this->params_.cost_rollout_dim_, this->stream_, false);
    // Copy the costs back to the host
    HANDLE_ERROR(cudaMemcpyAsync(this->trajectory_costs_.data(), this->trajectory_costs_d_,
                                 NUM_SAMPLES * sizeof(float), cudaMemcpyDeviceToHost, this->stream_));
    HANDLE_ERROR(cudaStreamSynchronize(this->stream_));
    
    // Setup vector to hold top k weight indices
    int top_k_to_keep = NUM_SAMPLES * this->params_.top_k_percentage;~\label{line:cem_elite_set_1}~
    calculateEliteSet(this->trajectory_costs_, NUM_SAMPLES, top_k_to_keep, top_k_indices_);~\label{line:cem_elite_set_2}~
    // keep weights of the elite set~\label{line:cem_elite_set_3}~
    float min_elite_value = this->trajectory_costs_[top_k_indices_.back()];~\label{line:cem_elite_set_4}~
    std::replace_if(~\label{line:cem_elite_set_5}~
        this->trajectory_costs_.data(), this->trajectory_costs_.data() + NUM_SAMPLES,~\label{line:cem_elite_set_6}~
        [min_elite_value](float cost) { return cost < min_elite_value; }, 0.0f);~\label{line:cem_elite_set_7}~

    // Copy weights back to device
    HANDLE_ERROR(cudaMemcpyAsync(this->trajectory_costs_d_, this->trajectory_costs_.data(),
                                 NUM_SAMPLES * sizeof(float), cudaMemcpyHostToDevice, this->stream_));
    // Compute the normalizer
    this->setNormalizer(mppi_common::computeNormalizer(this->trajectory_costs_.data(), NUM_SAMPLES));
    // Calculate the new mean
    this->sampler_->updateDistributionParamsFromDevice(this->trajectory_costs_d_, this->getNormalizerCost(),
                                                       0, false);
    // Transfer the new control back to the host and synchronize stream
    this->sampler_->setHostOptimalControlSequence(this->control_.data(), 0, true);
  }
  // Calculate optimal state and output trajectory from the current state and optimal control
  this->computeOutputTrajectoryHelper(this->output_, this->state_, state, this->control_);
}
\end{lstlisting}
\caption{Basic \acs{CEM} \lstinline{computeControl()} implementation. This method copies the initial state to the GPU, creates the control samples, calculates the state trajectories and costs of each sample, creates the elite set, updates the mean of the sampling distribution and copies that mean back as the optimal control sequence.}
\label{code:cem_controller_source}
\end{code*}
We walk through how to create a new sampling-based Controller using \ac{CEM} \cite{rubinstein1999cross} as our desired algorithm. 
Succinctly put, the \ac{CEM} method samples control trajectories from a Gaussian Distribution, and uses the best $k$ samples, known as the \emph{elite set}, to calculate new parameters for the Guassian distribution. 
For this example, we simplify it to have constant variance and the elite set is used to update the mean of the distribution. 
The major \ac{CEM} parameter is the size of the elite set, so we create a new parameter \lstinline{struct} in \cref{code:cem_controller_params} capturing this as a percentage of the total number of samples. 
From there, the methods that must be overwritten from the base Controller API are \lstinline{getControllerName()}, \lstinline{computeControl()}, and \lstinline{calculateSampledStateTrajectories()}.
Like in the other custom classes, \lstinline{getControllerName()} returns the name of the new controller, i.e. \ac{CEM}. 
Next, \lstinline{computeControl()} is the main method of the Controller class. 
It takes the new initial state and calculates the new optimal control sequence from that starting position; an example is shown in \cref{code:cem_controller_source}. 
The basic steps are to move the shifted optimal control sequence and initial state to the GPU, generate control samples, run the samples through the Dynamics and Cost function, calculate the weights of each sample, use these weights to update the parameters of the sampling distribution, get the new optimal control sequence, and calculate the corresponding optimal state trajectory. 
In \cref{line:cem_elite_set_1,line:cem_elite_set_2,line:cem_elite_set_3,line:cem_elite_set_4,line:cem_elite_set_5,line:cem_elite_set_6,line:cem_elite_set_7}, we add an additional method, \lstinline{calcualteEliteSet()}, to find the elite set and zero out the weights of every other sample specifically for \ac{CEM}. 
The final method to overwrite, \lstinline{calculateSampledStateTrajectories()}, is a method used to return a subset of sampled trajectories from the latest optimization round from the GPU to the CPU. 
Users can set a percentage of the sampled trajectories they would like returned with \lstinline{setPercentageSampledControlTrajectoriesHelper()} and this method will generate state trajectories for those samples so that they can then be used for visualization in programs such as RViz.
For this simple example, we have no visualization system to plug this data into so we leave this method empty for our \ac{CEM} implementation.

\subsubsection{Sampling Distributions}
\label{subsubsec:sampling_distributions}
The Sampling Distribution class is where the choice of how to sample the control distributions is conducted. 
Different sampling distributions can have significant impact to the controller performance and is still a large area of research being explored. 
The current sampling distributions implemented are Gaussian, Colored Noise \cite{vlahov2024low}, \ac{NLN} noise \cite{mohamed2022autonomous}, and Smooth-\ac{MPPI} \cite{kim2022smooth}, but even in those implementations, there are further options and tweaks that we found to improve performance in the past. 
Some of these include having a percentage of samples sampled from a zero-mean distribution rather than the previous optimal control sequence, using the mean with no noise as a sample, allowing for time-varying standard deviations, and the ability to disable the importance sampling weight.

The Sampling Distribution API leaves enough flexibility to allow for multi-hypothesis distributions such as \ac{GMM} distributions or Stein-Variational distributions \cite{wang2021variational,lambert2021stein}. In addition, the \lstinline{readControlSample()} method used to get the control sample for a particular sample, time, and system takes in the current output which allows for feedback-based sampling such as done in normalizing flow approaches \cite{power2022variational,sacks2023learning}. The essential methods to focus on when implementing a new Sampling Distribution are \lstinline{generateSamples()} and \lstinline{readControlSample()}.

\section{Benchmarks}
\label{sec:benchmarks}




In order to see the improvements our library can provide, we decided to compare against three other implementations of \ac{MPPI} publicly available. 
The first comparison is with the \ac{MPPI} implementation in AutoRally \cite{goldfain2021autorally}. 
This implementation was the starting point of our new library, MPPI-Generic, and so we want to compare to see how well we can perform to our predecessor. 
The Autorally implementation is written in C++/CUDA, is compatible with \ac{ROS}, features multiple dynamics models including linear basis functions, simple kinematics, and \ac{NN}-based models focused on the Autorally hardware platform. 
There is only one Cost Function available but it makes use of CUDA textures querying into an obstacle map. 
Additionally, it has been shown to run in real-time on hardware to great success \cite{goldfain2019autorally,williams2016aggressive,williams2018information}. 
However, the Autorally implementation is written for use on the Autorally platform and  has no general Cost Function, Dynamics, or Sampling Distribution APIs to extend. In order to use it for different problems such as flying a quadrotor, the \ac{MPPI} implementation would need significant modification. 

The next implementation we compare against is Nav2's \ac{MPPI}. 
As of \ac{ROS} Humble, there is a CPU implementation of \ac{MPPI} in the \ac{ROS} navigation stack \cite{macenski2023desks}; for our testing, we utilized the version packaged for ROS 2 Iron.
This CPU implementation \cite{budyakov2023ros2} is written in C++ and looks to make heavy use of AVX or other vectorized instructions to improve performance. 
There is a small selection of dynamics models (Differential Drive, Ackermann, and Omni-directional) and cost functions that are focused around wheeled robots navigating through obstacle-laden environments. 
It allows users to implement new cost functions as critics through a plugin system and new dynamics models based off its MotionModel base class.
This implementation will only become more widespread as ROS2 adoption continues to grow over the coming years, making it an essential benchmark. 
Unfortunately, it does have some drawbacks as it  has no implementation of Tube-\ac{MPPI} or \ac{RMPPI}, and is only available in ROS2 Humble or newer.
This means that it might not be usable on existing hardware platforms that are unable to upgrade their systems.

The last implementation of \ac{MPPI} we compare against is in TorchRL \cite{bou2024torchrl}. 
TorchRL is an open-source \ac{RL} Python library written by Meta AI, the developers of PyTorch itself. 
As such, it is widely trusted and available to researchers who are already familiar with PyTorch and Python. 
The TorchRL implementation works on both CPUs and GPUs and allows for custom dynamics and cost functions through the extension of base Environment class \cite{torchrlcontributors2023mppi}. 
However, while it does have GPU support, it is limited to the functionality of PyTorch meaning that there is no option to use CUDA textures to improve map queries or any direct control of shared memory usage on the GPU. 
In addition, being written in Python makes it fairly legible and easy to extend but can come at the cost of performance when compared to C++ implementations. 

In order to compare our library against these three implementations, we recreated the same dynamics and cost function for each version of \ac{MPPI}. 
We chose the Differential Drive dynamics model and some of the cost function components that already exist in the Nav2 impelemtation as the baseline. 
We used the goal position quadratic cost, goal angle quadratic cost, and the costmap-based obstacle cost components so that we could maintain a fairly simple cost function that allows us to show the capabilities of our library. 
We implemented these dynamics and cost functions in both CUDA and Python. 
The CUDA implementations were extensions of our base Dynamics and Cost Function APIs. 
We decided to use the same code in the Autorally implementation as well which required some minor rewriting to account for different method names and state dimensions. 
The Python implementation was an extension of the TorchRL base Environment class, and used PyTorch's JIT compiler to speed up performance when used in the TorchRL implementation. 
We used the same parameters for sampling, dynamics, cost function tuning, and MPPI hyperparameters across all implementations, summarized in \cref{tab:algorithm_parameters}.
\begin{table}[hb!]
    \centering
    \DeclareSIUnit[]\cell{\text{cell}}
    \caption{Algorithm Parameters}
    \label{tab:algorithm_parameters}
    \begin{tabular}{cc}
        \toprule
        \text{Parameter} & \text{Value} \\
        \midrule
        dt & 0.02 s \\
        wheel radius & 1.0 \si{m} \\ 
        wheel length & 1.0 \si{m} \\
        max velocity & 0.5 \si{\meter/\second} \\
        min velocity & -0.35 \si{\meter/\second} \\
        min rotation & -0.5 \si{\radian/\second} \\
        max rotation & 0.5 \si{\radian/\second} \\
        \toprule
        \multicolumn{2}{c}{MPPI Parameters} \\
        \midrule
        $\lambda$ & 1.0 \\
        control standard deviation & 0.2 \\
        MPC Horizon & 100 timesteps \\
        \toprule
        \multicolumn{2}{c}{Cost Parameters} \\
        \midrule
        Dist. to goal coefficient & 5 \\
        Angular Dist. to goal coeff & 5 \\
        Obstacle Cost & 20 \\
        Map width & 11 \si{\meter} \\
        Map Height & 11 \si{\meter} \\
        Map Resolution & 0.1 \si{\m/\cell}\\
        \bottomrule
    \end{tabular}
\end{table}

For both Autorally and MPPI-Generic, there are further performance-enhancing options available such as block size choice. 
We ended up using the same block sizes for both Autorally and MPPI-Generic across all tests, shown in \cref{tab:gpu_parameter}. 
\begin{table}[hb!]
    \centering
    \caption{GPU Performance Choices}
    \label{tab:gpu_parameter}
    \begin{tabular}{cc}
        \toprule
        \text{Parameter} & \text{Value} \\
        \midrule
        Dynamics thread block x dim. & 64 \\
        Dynamics thread block y dim. & 4 \\
        Cost thread block x dim. & 64 \\
        Cost thread block y dim. & 1 \\
        \bottomrule
    \end{tabular}
\end{table}
As a result, the optimization times shown are not going to be the fastest possible performance that can be achieved on any given GPU but these tests should still serve as a useful benchmark to understand the average performance that can be achieved.
Our test was timing how long each implementation of \ac{MPPI} would take to return an optimal control sequence when provided an initial state, $\vb{x}_0$.
We ran each of the Autorally, MPPI-Generic, and Nav2 implementations $10,000$ times to produce optimal trajectories with 128, 256, 512, 1024, 2048, 4096, 6144, 8192, and 16,384 samples; the TorchRL implementation was only run $1000$ times due to it being too slow to compute, even when using the GPU.
The comparisons were run across a variety of hardware including a Jetson Nano to see what bottlenecks each implementation might have. 
The Jetson Nano was unfortunately only able to run the MPPI-Generic and Autorally \ac{MPPI} implementations as the last supported PyTorch version and the lastest TorchRL libraries were incompatible, and the Nav2 implementation was unable to compile.
GPUs tested ranged from a NVIDIA GTX 1050 Ti to a NVIDIA RTX 4090. 
Most tests were performed on an Intel 13900K which is one of the fastest available CPUs at the time of this writing in order to prevent the CPU being the bottleneck for the mostly GPU-based comparison\ifbool{arxivVersion}{; however, we also ran tests on an AMD Ryzen 5 5600x to see the difference in performance on a lower-end CPU.
The \ac{MPPI} optimization times across all hardware can be seen in \cref{tab:experiment_summary}.
}{.}
The code used to do these comparisons is available at \url{https://github.com/ACDSLab/MPPI_Paper_Example_Code}.

\subsection{Results}
\label{subsec:results}
\begin{figure}[th!]
    \centering
    \includegraphics[width=0.8\linewidth]{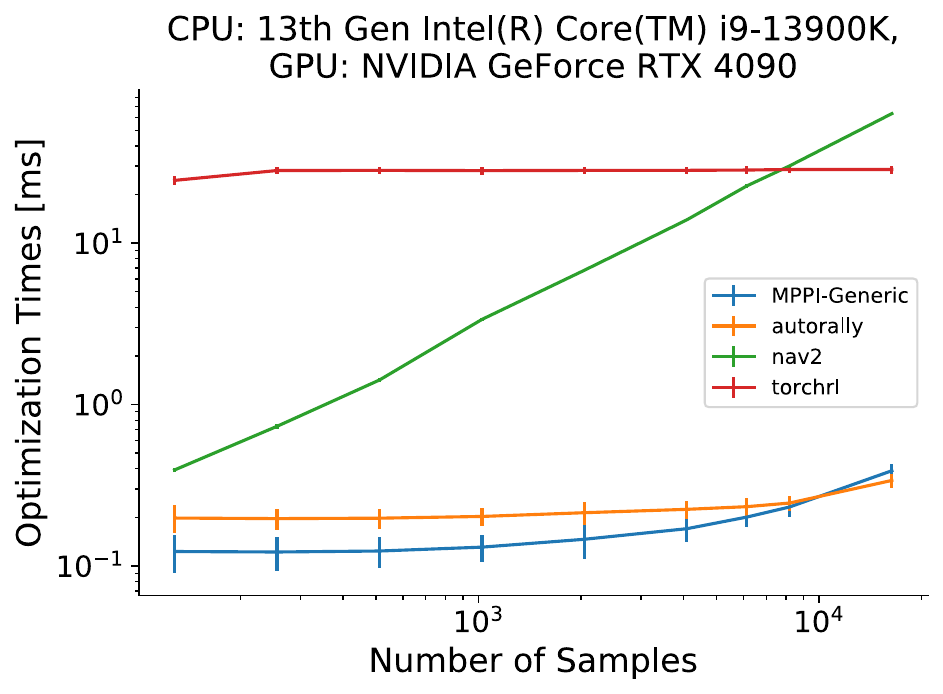}
    \caption{Optimization times for all MPPI implementations on a hardware system with a RTX 4090 and an Intel 13900K over a variety of number of samples. The Nav2 CPU implementation grows linearly as the number of samples increase while GPU implementations grow more slowly.}
    \label{fig:4090_results}
\end{figure}
While this results section focuses on the computational speed of our algorithm for fairly simple dynamics and cost functions, readers may be curious to know how this library performs in real-world applications. Adding those results to this paper would end up requiring too much space for this publication; instead, we point readers to other papers where this code base has been used in more complex scenarios \cite{vlahov2024low,patel2023modelpredictive,gandhi2021robust,gandhi2022safety,gibson2023multistep}.

\begin{figure}
    \centering
    \includegraphics[width=0.8\linewidth]{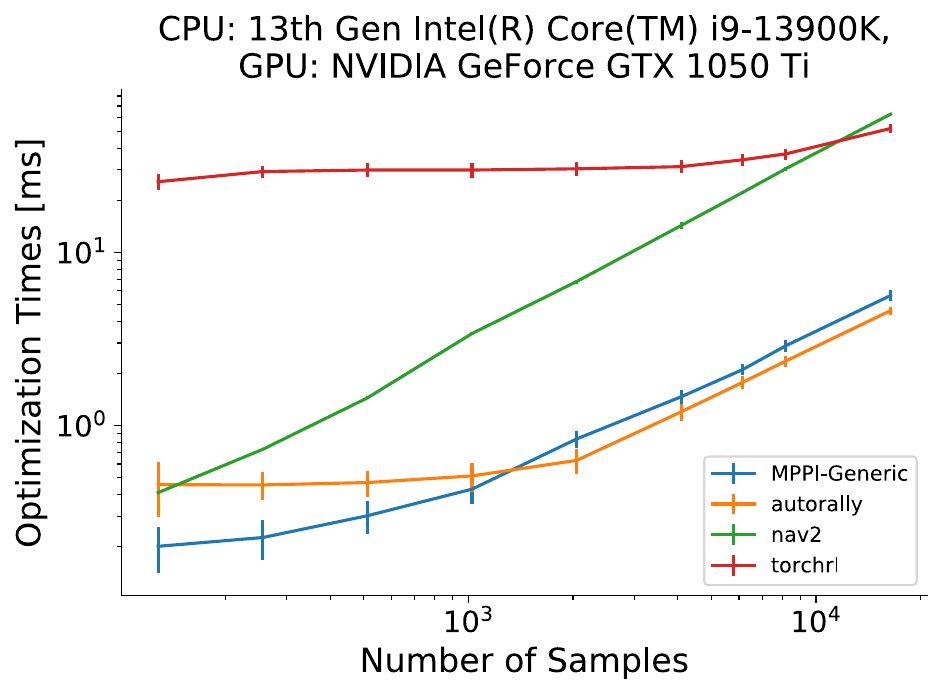}
    \caption{Optimization times for all MPPI implementations on a hardware system with a GTX 1050 Ti and an Intel 13900K over a variety of number of samples. MPPI-Generic and AutoRally on this older hardware eventually start to scale linearly with the number of samples but does so at a much lower rate with our library compared to Nav2 or TorchRL.}
    \label{fig:1050_results}
\end{figure}

\ifbool{arxivVersion}{Going over all of the collected data would take too much room for this paper so we shall instead try to pull out interesting highlights to discuss. 
Full results can be seen in \cref{tab:experiment_summary}.}{}
First, we look at the results on the most powerful system tested, using an Intel i9-13900K and an NVIDIA RTX 4090 in \cref{fig:4090_results}. 
As the number of samples increase, the CPU-bound Nav2 method increases in optimization times in a linear fashion immediately.
Every other method uses the GPU which can provide headroom to increase the number of samples as the GPU is not fully utilized at small sample sizes.
As we hit $16,384$ samples, the AutoRally implementation starts to have lower optimization times than \ac{MPPI}-Generic. We will see this trend continue in \cref{fig:1050_results}. 

When looking at older and lower-end NVIDIA hardware such as the GTX 1050 Ti, our library still performs well compared to other implementations as seen in \cref{fig:1050_results}. 
Only when the number of samples is at 128 does the Nav2 implementation on an Intel 13900k match the performance of the AutoRally implementation on this older GPU. 
However, we can see that Nav2 implementation again outperforms the TorchRL implementation until we hit 16,384 samples.
MPPI-Generic is still more performant at these lower number of samples and eventually it scales linearly as we get to thousands of samples. 
The TorchRL implementation also finally starts to show some GPU bottle-necking as we start to see optimization times increasing as we reach over 6144 samples.
There is also a moment where the MPPI-Generic library optimization time grows to be larger than the AutoRally implementation. 
That occurs when we switch from using the split kernels (\cref{subsec:split-kernel}) to the combined kernel (\cref{subsec:combined-kernel}). The AutoRally implementation uses a combined kernel with fewer GPU synchronization points due to strictly requiring forward Euler integration for the dynamics. 
At the small hit to performance in the combined kernel, our library allows for many more features, such as multi-threaded cost functions, use of shared memory in the cost function, and implementation of more computationally-heavy integration methods such as Runge-Kutta or backward Euler integration. 
And while we see a hit to performance when using the combined kernel compared to AutoRally, we still see that the split kernel is faster for up to 2048 samples.

\begin{figure}
    \centering
    \includegraphics[width=0.8\linewidth]{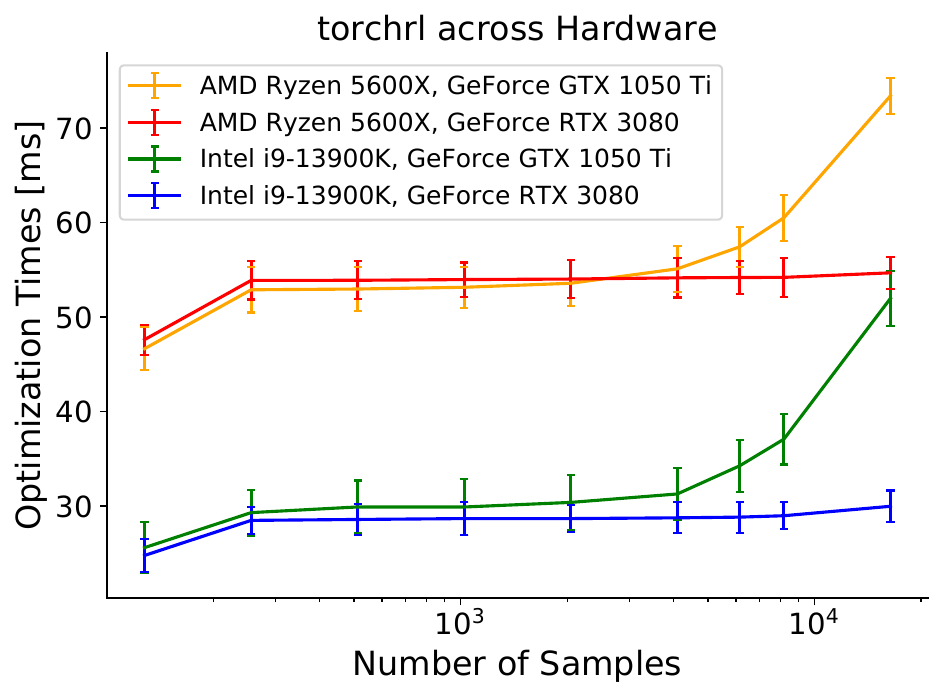}
    \caption{Optimization times for the TorchRL implementation across different CPUs and GPUs. TorchRL computation times are more dependent on the CPU as the RTX 3080 with an AMD 5600X ends up slower than a GTX 1050 Ti with an Intel 13900k.}
    \label{fig:torchrl_results}
\end{figure}

The TorchRL implementation is notably performing quite poorly in \cref{fig:4090_results,fig:1050_results} with runtimes being around $28 \si{\ms}$ no matter the number of samples. 
Looking at TorchRL-specific results in \cref{fig:torchrl_results}, the TorchRL implementation seems to be heavily CPU-bound. 
A low-end GPU (1050 Ti) combined with a high-end CPU (Intel 13900K) can achieve better optimization times than a low-end CPU (AMD 5600X) combined with a high-end GPU (RTX 3080). 

\begin{figure}
    \centering
    \includegraphics[width=0.8\linewidth]{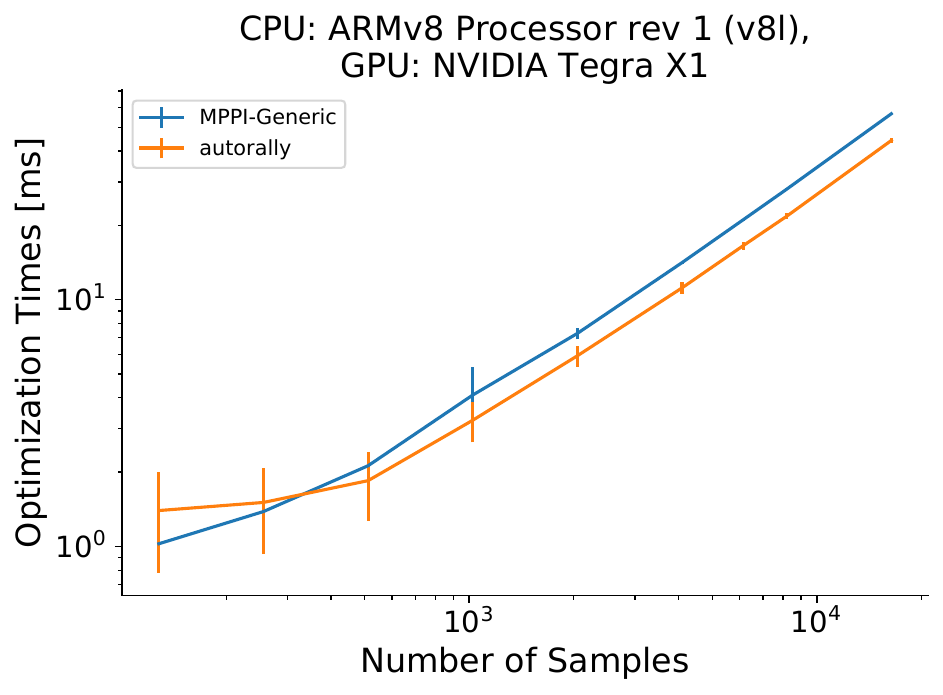}
    \caption{Optimization times for MPPI implementations on a Jetson Nano over a variety of number of samples. MPPI-Generic and AutoRally on this low-power hardware can still achieve sub-$10 \si{\ms}$ optimization times for even 2048 samples. The AutoRally implementation quickly surpasses our implementation in optimization times.}
    \label{fig:jetson_results}
\end{figure}

We also conducted tests on a 2019 Jetson Nano to show that even on relatively low-power and older systems, our library can still be used. 
As the latest version of CUDA supporting the Jetson Nano is 10.2 and the OS is Ubuntu 18.04, both the TorchRL and Nav2 \ac{MPPI} implementations were not compatible. 
As such, we only have results comparing our MPPI-Generic implementation to the AutoRally implementation in \cref{fig:jetson_results}.
Here, the AutoRally implementation starts having faster compute times around 512 samples. Again, this is due to our library switching to the combined kernel which will be slower. 
However, our library on a Jetson Nano at 2048 samples has a roughly equivalent computation time to that of 2048 samples of the Nav2 implementation on the Intel 13900K process, showing that our GPU parallelization can allow for real-time optimization even on portable systems.

\begin{figure}
    \centering
    \includegraphics[width=0.8\linewidth]{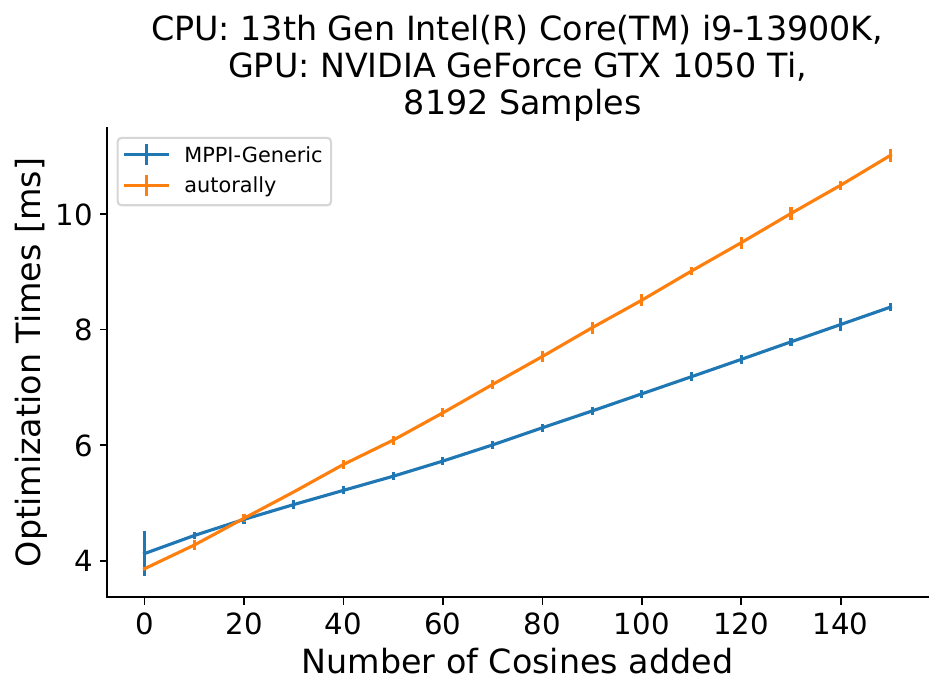}
    \caption{Optimization Times for \ac{MPPI}-Generic and AutoRally implementations as the computation time of the cost function increases. Using an Intel 13900K, NVIDIA GTX 1050 Ti, and 8192 samples, our library implementation starts to outperform the AutoRally implementation when 20+ sequential cosine operations are added to the cost function.}
    \label{fig:cost_complexity_results}
\end{figure}

In addition, we see the benefits of our library as we increase the computation time of the cost function. 
At this point, the TorchRL and Nav2 implementations have been shown to be slow in comparison to the other implementations and are thus dropped from this cost complexity comparison.
\begin{code}[ht!]
\begin{lstlisting}[language=C++,firstline=1,lastline=15,escapechar=~]
float computeStateCost(...) {
  float cost = 1.0;
  for (int i = 0; i < NUM_COSINES; i++) {
    cost = cos(cost);
  }
  cost *= 0.0;
  // Continue to regular cost function
  ...
}
\end{lstlisting}
\caption{Computation time inflation code added to the cost function. We add a configurable amount of calls to \lstinline{cos()} as this is a computationally heavy function to run.}
\label{code:cost_time_inflation}
\end{code}
We artificially inflate the computation time of the cost function with \cref{code:cost_time_inflation} to judge how well the implementations scale to more complex cost functions. 
In \cref{fig:cost_complexity_results}, we see how increasing the computation time of the cost function scales for both implementations over the same hardware and for the same number of samples.

\subsection{Comparisons to sampling-efficient algorithms}
\label{subsec:dmd_mpc_comparisons}
\begin{figure}[htp!]
    \centering
    \includegraphics[width=0.9\linewidth]{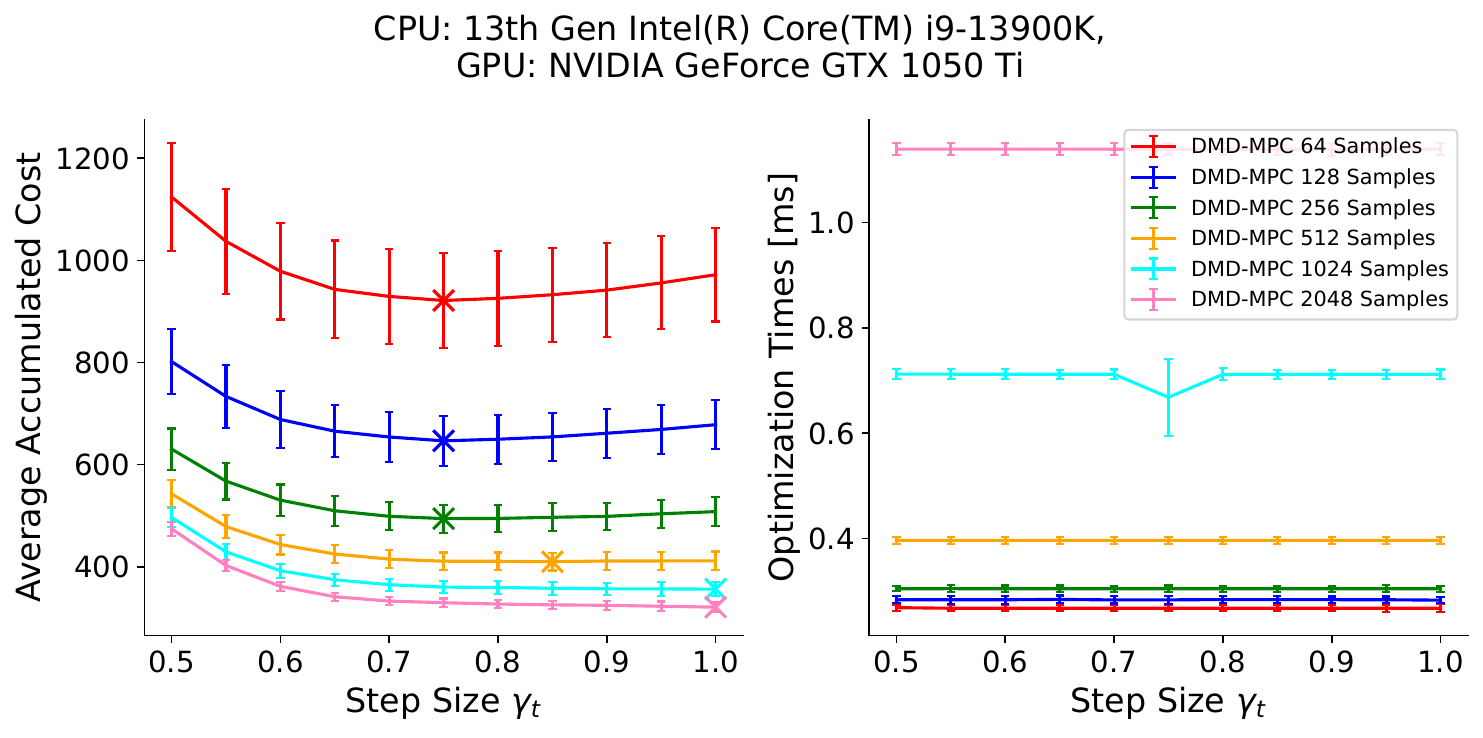}
    \caption{Average Accumulated Costs (left) and Optimization Times (right) with error bars signifying one standard deviation for a variety of step sizes and number of samples for \acs{DMD-MPC}. $\times$ indicates the step size that achieves the lowest cost for a given number of samples. When using a low number of samples, a lower \acs{DMD-MPC} step size provides the lowest average cost. However, as the number of samples increase, the best step size choice becomes $\gamma_t = 1.0$ which is equivalent to the normal \ac{MPPI} update law. With our library, increasing the number of samples to the point where the step size is no longer useful is still able to be run at over 800 Hz on the NVIDIA GTX 1050 Ti.}
    \label{fig:dmd-results}
\end{figure}

While we have shown that our implementation of \ac{MPPI} can have faster computation times and a lot of flexibility in applications, there remains a question of how to balance between the number of samples and real-time performance. 
Our work decreases the computation time for sampling which in turn allows more samples in the same computation time, while other works have tried to reduce the amount of samples needed to evaluate the optimal control trajectory.
Many authors \cite{vlahov2024low,mohamed2022autonomous,honda2023stein,kim2022smooth,power2022variational,sacks2023learning} have tried to do this by changing the sampling distribution; depending on the derivation of \ac{MPPI}, different sampling distributions can be considered to have the same update rule.
In \cite{wagener2019online}, the authors introduce a new update rule through a generalization of \ac{MPC} algorithms called \ac{DMD-MPC} defined by the choice of shaping function, $S(\cdot)$, Sampling Distribution $\pi_\theta$, and Bregman Divergence $D_\Psi(\cdot, \cdot)$ which determines how close the new optimal control trajectory should remain to the previous. Using the exponential function, Gaussian sampling, and the \ac{KL} Divergence, they derive a modification to the \ac{MPPI} update law in \cref{eq:mppi_update_rule} that introduces a step size parameter $\gamma_t \geq 0$:
\begin{align}
    \vb{u}_{t}^{k+1} = \PP{1 - \gamma_t} \vb{u}_{t}^k + \gamma_t \frac{\Expectation[V \sim \pi_\theta]{\expf{-\frac{1}{\lambda}\mathcal{J}\PP{V}}\vb{v}_t}}{\Expectation[V \sim \pi_\theta]{\expf{-\frac{1}{\lambda}\mathcal{J}\PP{V}}}},
\end{align}
where $\vb{v}_t$ is the control value at time $t$ from the sampled control sequence $V$, and $\mathcal{J}$ is as we defined in \cref{eq:cost_trajectory_shorthand}. 
In their results, the addition of a step size can improve performance when using a low number of samples. 
However, once the number of samples increases beyond a certain point, the optimal step size ends up being $1.0$ which is equivalent to the original \ac{MPPI} update law. 
Having this option is useful in cases where you have a low computational budget. 
We show as long as you have a NVIDIA GPU from the last decade, you have enough computational budget to use more samples without needing to tune a step size. 

Looking at \cref{fig:dmd-results}, we ran a 2D double integrator system with state $[x,y,v_x,v_y]$ and control $[a_x, a_y]$ with the cost shown in \cref{eq:double_integrator_circle_cost}:
\begin{align}
\begin{split}
    J &= 1000 \PP{\indicator_{\left\{\PP{x^2+y^2} \leq 1.875^2\right\}} + \indicator_{\left\{\PP{x^2+y^2} \geq 2.125^2\right\}}} \\
    &+ 2\abs{2 - \sqrt{v_x^2 + v_y^2}} + 2\abs{4 - \PP{xv_y - yv_x}}.
    \label{eq:double_integrator_circle_cost}
\end{split}
\end{align}
This cost function heavily penalizes the system from leaving a circle of radius $2\si{\m}$ with width $0.125\si{\m}$, has an $L_1$ cost on speed to maintain $2\si{\m\per\s}$, and has an $L_1$ cost on the angular momentum being close to $4\si{\m^2\per\s}$. This all combines to encourage the system to move around the circle allowing some small deviation from the center line in a clockwise manner.
This system was simulated for 1000 timesteps and the cost was accumulated over that period. This simulation was run 1000 times to ensure consistent cost evaluations. 
As the number of samples increase, the optimal step size (marked with a x) increases to $1.0$. 
In addition, the computation time increase for using a number of samples where the step size is irrelevant is minimal (an increase of about $0.4\si{\ms}$).
There is also headroom to increase the complexity of the dynamics and cost function and still run the controller at over $100 \si{Hz}$.

\section{Conclusion}
\label{sec:conclusion}

In this paper, we introduce a new sampling-based optimization C++/CUDA library called \ac{MPPI}-Generic. 
It contains implementations of \ac{MPPI}, Tube-\ac{MPPI}, and \ac{RMPPI} controllers as well as an API that allows these controllers to be used with multiple dynamics and cost functions. 
We went through various ways that researchers could use this library, from using pre-defined dynamics and cost functions to implementing new sampling-based \ac{MPC} controllers. 
We discussed the methods used to improve the computational performance and conducted performances comparisons against other widely-available implementations of \ac{MPPI} over a variety of computer hardware to show the performance benefits our library can provide. 
Finally, we compared against a sample-efficient form of \ac{MPPI} to show that with the speed improvements of our library, using more samples is a viable alternative with little hit to computation times. 
We plan to keep working on the library to add more capabilities and usage improvements such as a Python wrapper in the future.

\section{Acknowledgements}
This research was funded in part by the Office of Naval Research under contract N00014-21-1-2074, Defense Advanced Research Projects Agency (DARPA), Jet Propulsion Laboratory (JPL), California Institute of Technology, under contract 80NM0018D0004 with the National Aeronautics and Space Administration, NASA LaRC, and Sandia National Laboratories, a multimission laboratory managed and operated by National Technology and Engineering Solutions of Sandia, LLC., a wholly owned subsidiary of Honeywell International, Inc., for the U.S. Department of Energy’s National Nuclear Security Administration under contract DE-NA-0003525. The views, opinions and/or findings expressed are those of the author and should not be interpreted as representing the official views or policies of the Department of Defense or the U.S. Government.

\ifbool{arxivVersion}{\input{results_table}}{}

\ifbool{biblatexReference}{
\printbibliography
\printbibliography[type=software,heading=subbibliography,title={Software Sources}]
}{
\balance
\bibliographystyle{IEEEtran}
\bibliography{references}
}

\ifbool{arxivVersion}{}{
\begin{IEEEbiography}[{
\includegraphics[width=1in,clip,trim=0 0 0 0]
{Figures/Authors/bogdan_cropped.jpg}}]{Bogdan Vlahov}
received the BS degree in Computer Science and the BE degree in Mechanical Engineering from Purdue University in 2017. He is currently working towards a PhD degree in Robotics from Georgia Institute of Technology. His research interests include high performance computing for robotics, nonlinear control theory and optimization, and artificial intelligence.
\end{IEEEbiography}
\begin{IEEEbiography}[{\includegraphics[width=1in,clip]{Figures/Authors/jason.jpeg}}]{Jason Gibson}
received a BS in Computer Science from the Georgia Institute of Technology in 2019 and is currently working on a PhD in Robotics from the same institution. His research centers around combining learning with dynamics modeling and control for various autonomous platforms. These models are used for risk-aware planning with perceptual inputs on hardware platforms.
\end{IEEEbiography}
\begin{IEEEbiography}[{\includegraphics[width=1in,clip]{Figures/Authors/manan_cropped.png}}]{Manan Gandhi}
received a PhD in Aerospace Engineering from the Georgia Institute of Technology in 2023 and is currently a Senior Member of Technical Staff at Sandia National Laboratories, with expertise in stochastic optimal control, safe control for high consequence systems, and model predictive control. Currently he is working on multiagent coordination and navigation, with a side interest in biological modeling and optimal control of cardiopulmonary resuscitation.
\end{IEEEbiography}
\begin{IEEEbiography}[{\includegraphics[width=1in,clip,trim=300 100 250 155]{Figures/Authors/Theodorou.jpeg}}]{Evangelos Theodorou} is an Associate Professor with the School of Aerospace Engineering at Georgia Institute of Technology and is the Director of the Autonomous Control and Decision Systems Laboratory. 
Prof. Theodorou is an affiliate of the Institute of Robotics and Intelligent Machines (IRIM) and the Centers for Machine Learning Research. 
His research interest spans the areas of Control Systems, Optimization, Autonomy,
Robotics.
\end{IEEEbiography}
}

\end{document}

%% file: packages.tex
\usepackage{amsfonts}       
\usepackage{amssymb}
\usepackage{amsthm}
\usepackage{balance}
\usepackage{booktabs} 
\usepackage{graphicx}
\usepackage{enumitem}
\usepackage{float}
\usepackage{listings}
\usepackage{supertabular}
\usepackage{mathtools}
\usepackage{mathrsfs}
\usepackage{multicol}
\usepackage{newfloat}
\usepackage{siunitx}
\usepackage{svg}
\usepackage[dvipsnames]{xcolor}
\usepackage[ruled,vlined]{algorithm2e}

\usepackage{etoolbox}
\newbool{biblatexReference}

\newbool{arxivVersion}

\booltrue{arxivVersion} 

\ifbool{biblatexReference}{
}{
\usepackage{usebib}

\ifbool{arxivVersion}{
    \usepackage[backref=page]{hyperref}
}{
    \usepackage{hyperref}
}
}

\hypersetup{
    colorlinks=true,
    linkcolor=OliveGreen,
    filecolor=magenta,      
    urlcolor=blue,
    citecolor=purple,
}
\usepackage[capitalise]{cleveref}
\usepackage[caption=false, font=footnotesize]{subfig}

\usepackage[nolist]{acronym}

\definecolor{listinggray}{gray}{0.9}
\definecolor{lbcolor}{rgb}{0.9,0.9,0.9}
\definecolor{CommentColor}{rgb}{0,0.4,0}
\definecolor{LineNumberColor}{rgb}{0.2,0.15,0.75}
\definecolor{KeywordColor}{rgb}{0,0,1}
\definecolor{StringColor}{rgb}{0.627,0.126,0.941}
\makeatletter%
\def\lst@PlaceNumber{\makebox[\dimexpr 1em+\lst@numbersep][l]{\normalfont
  \lst@numberstyle{\thelstnumber}}}%
\makeatother%

\DeclareGraphicsExtensions{.png,.pdf}
\makeatletter 
\usepackage[normalem]{ulem}

%% file: commands.tex
\Crefname{code}{Lst.}{Lst.}

\DeclareFloatingEnvironment[
    fileext=locode,
    listname={List of Listings},
    name=Listing,
    placement=tbhp,
]{code}

\newcommand{\abs}[1]{\left|#1\right|}

\newcommand{\vb}[1]{{\bf #1}}

\newcommand{\R}{\mathbb{R}}
\newcommand{\PP}[1]{\left(#1\right)}
\newcommand{\PB}[1]{\left[#1\right]}
\newcommand{\PCB}[1]{\left\{#1\right\}}

\newcommand{\expf}[1]{\exp\left(#1\right)}
\newcommand{\Expectation}[2][]{\mathbb{E}_{#1}\left[#2\right]}
\newcommand{\KL}[2]{\mathbb{KL}\left(#1\ ||\ #2 \right)}

\newcommand{\normal}[1]{\mathcal{N}\PP{#1}}
\newcommand*{\indicator}{\text{\usefont{U}{bbold}{m}{n}1}}

\DeclareMathOperator*{\argmax}{arg\,max}
\DeclareMathOperator*{\argmin}{arg\,min}

\DeclareMathOperator{\J}{\bf J}

\theoremstyle{definition}

\begin{acronym}
\acro{ADMM}{Alternating Direction Method of Multipliers}
\acro{AR}{AutoRally}
\acro{CEM}{Cross-Entropy Method}
\acro{CNN}{Convolutional Neural Network}
\acro{ConvLSTM}{Convolutional Long Short Term Memory}
\acro{CRTP}{Curiously Recurring Template Pattern}
\acro{CTG}{Cost-To-Go}
\acro{CVaR}{Conditional Value-at-Risk}
\acro{DDP}{Differential Dynamic Programming}
\acro{DMD-MPC}{Dynamic Mirror Descent Model Predictive Control}
\acro{EKF}{Extended Kalman Filter}
\acro{EM}{Expectation Maximization}
\acro{ES}{Evolutionary Strategies}
\acro{GMM}{Gaussian Mixture Model}
\acro{GP}{Gaussian Process}
\acro{GRU}{Gated Recurrent Unit}
\acro{FT}{Fourier Transform}
\acro{FCNN}{Fully-Connected Neural Network}
\acro{iCEM}{improved Cross-Entropy Method}
\acro{iFFT}{inverse Fast Fourier Transform}
\acro{iLQR}{iterative Linear Quadratic Regulator}
\acro{IS}{Importance Sampler}
\acro{KL}{Kullback Leibler}
\acro{LSTM}{Long Short Term Memory}
\acro{MOO}{Multi-Objective Optimization}
\acro{MBRL}{Model-Based Reinforcement Learning}
\acro{MPC}{Model Predictive Control}
\acro{MPPI}{Model Predictive Path Integral}
\acro{NLN}{Normal log-Normal}
\acro{NN}{Neural Network}
\acro{PCA}{Principle Component Analysis}
\acro{PDF}{Probability Density Function}
\acro{PSD}{Power Spectral Density}
\acro{RCNN}{Region-based Convolutional Neural Network}
\acro{RL}{Reinforcement Learning}
\acro{ROS}{Robot Operating System}
\acro{RMPPI}{Robust-MPPI}
\acro{RPN}{Region Proposal Network}
\acro{RRT}{Rapidly Exploring Random Tree}
\acro{SQP}{Sequential Quadratic Programming}
\acro{VaR}{Value-at-Risk}
\acro{YOLO}{You Only Look Once}
\end{acronym}

\newcommand*{\pe}{\mathrel{+}=}


%% file: info_theory_short.tex
We shall follow the information-theoretic derivation from \cite{williams2018information} with some changes. In \cite{williams2018information}, the running cost definition was only based on the state trajectory and the authors showed how a control cost would emerge naturally from the importance sampling scheme. However, the derivation does not require this limited running cost definition, so we generalize our running cost to $\ell(\vb{x}_t,\vb{u}_t)$. We use the following shorthand to denote the cost for a control trajectory $U$ applied to the starting state, $\vb{x}_0$,
\begin{align}
\begin{split}
    \mathcal{J}\PP{U} &:= \vb{J}(X, U) \\
    \text{s.t. } &\vb{x}_{t+1} = \vb{F}\PP{\vb{x}_t, \vb{u}_t}
\end{split}
\label{eq:cost_trajectory_shorthand}
\end{align}
Our goal is to find a control trajectory, $U^*$, that minimizes \cref{eq:nonlinear_control_optimization} through the use of sampling.
We start by defining the \emph{free energy} of our system as
\begin{align}
    \mathcal{F}\PP{\mathcal{J}, \mathbb{P}, \lambda;\vb{x}_0} &= -\lambda\ln\PP{\Expectation[V \sim \mathbb{P}]{\exp\PP{-\frac{1}{\lambda} \mathcal{J}\PP{V}}}} \label{eq:free_energy}
\end{align}
where $\vb{x}_0$ is the initial state, $\lambda \in \R^+$ is the inverse temperature, and $V$ is the control trajectory sampled from some base distribution $\mathbb{P}$ with \ac{PDF} $p$. 
The base distribution $\mathbb{P}$ can be any distribution, making it potentially computationally intractable.
Introducing a new distribution $\mathbb{Q}$ that is absolutely continuous with $\mathbb{P}$, we use importance sampling and Jensen's Inequality to get the following,
\begin{align}
    \mathcal{F}\PP{\mathcal{J}, \mathbb{P}, \lambda;\vb{x}_0} &\leq \Expectation[V \sim \mathbb{Q}]{\mathcal{J}\PP{V}} + \lambda \KL{\mathbb{Q}}{\mathbb{P}} \label{eq:free_energy_bound}
\end{align}
where $\KL{\cdot}{\cdot}$ is the \ac{KL} divergence between $\mathbb{Q}$ and $\mathbb{P}$.
This introduces an upper bound on our free energy when we sample from $\mathbb{Q}$ instead of the original $\mathbb{P}$; this upper bound becomes a strict equality if $\mathbb{Q}^*$ has the following density: 
\begin{align}
    q^*(V) &= \frac{1}{\eta} \exp\PP{-\frac{1}{\lambda}\mathcal{J}\PP{V}}p(V) \label{eq:optimal_distribution}\\
    \eta &= \Expectation[\hat{V} \sim \mathbb{P}]{\exp\PP{-\frac{1}{\lambda}\mathcal{J}(\hat{V})}}
\end{align}

While $\mathbb{Q}^*$ is an optimal distribution that minimizes the free energy bound, the optimal \ac{PDF}, $q^*(V)$, still relies on the \ac{PDF} of the original distribution $\mathbb{P}$, meaning it still might be computationally intractable. 
Thus, we introduce a third distribution $\mathbb{S}_\theta$ that has controllable parameters $\theta$ and minimizes the \ac{KL} divergence between $\mathbb{S}_\theta$ and $\mathbb{Q}^*$,
\begin{align}
    \theta^* &= \argmin_{\theta \in \Theta} \KL{\mathbb{Q}^*}{\mathbb{S}_\theta} \\
    &=\argmax_{\theta \in \Theta} \Expectation[V \sim \mathbb{Q}^*]{\ln\PP{s\PP{V | \theta}}}. \label{eq:parameter_optimization}
\end{align}

\cref{eq:parameter_optimization} gives us an update rule for any probability distribution with parameters to match the optimal distribution. 
Leaving this aside for the moment, we show how to calculate the optimal control sequence by sampling from $\mathbb{S}_\theta$. From there, we will show how the update rule and the optimal control sequence can relate to each other depending on the choice of distributional family, $\mathbb{S}_\theta$. For the optimal control to use at time $t$, we shall look to the mean of the optimal distribution $\mathbb{Q}^*$,
\begin{align}
    \vb{u}_t^* &= \Expectation[V \sim \mathbb{Q}^*]{\vb{v}_t} \\
    &= \int_{V \sim\;\mathcal{U}} \vb{v}_t q^*\PP{V} dV
\end{align}
Using importance sampling to sample from $\mathbb{S}_\theta$ as well as \cref{eq:optimal_distribution} gives the following,
\begin{align}
    \vb{u}_t^* &= \int_{V \sim\;\mathcal{U}} \vb{v}_t \frac{q^*\PP{V}}{s\PP{V|\theta}} s\PP{V|\theta}dV \\
    &= \Expectation[V \sim \mathbb{S}_\theta]{w(V) \vb{v}_t} \label{eq:info_theory_opt_control_seq}\\
    w(V) &= \frac{1}{\eta}\exp\PP{-\frac{1}{\lambda} \mathcal{J}\PP{V}} \frac{p\PP{V}}{s\PP{V|\theta}}
    \label{eq:info_theory_weighting}
\end{align}
The only term in $w(V)$ that is unaccounted for is the importance sampling weight of $\mathbb{P}$ and $\mathbb{S}_\theta$. 

In \cite{williams2018information}, the base and parameterized distributions were both assumed to be Gaussian distributions with constant variance across time, $\Sigma$. The mean trajectory of $\mathbb{P}$ was assumed to be $\vb{0}_{n_u \times T}$ while the mean trajectory of $\mathbb{S}_\theta$ was the parameter to optimize, i.e. $\theta = \vb{U}^* = \PCB{\vb{u}^*_t}_{t=0}^{T-1}$. This led to the specific form of
\begin{align}
    w(V) &= \frac{1}{\eta}\exp\PP{-\frac{1}{\lambda}\mathcal{J}(V) - \sum_{t=0}^{T-1} \big(\vb{v}_t - \frac{1}{2} \vb{u}^*_t\big)^\top \Sigma^{-1} \vb{u}^*_t}
    \label{eq:information_theoretic_weighting}
\end{align}
and also meant that the solutions to \cref{eq:parameter_optimization,eq:info_theory_opt_control_seq} were the same.

%% file: results_table.tex
\begin{table*}[hb!]
\centering

\caption{MPPI Method Optimization Times at various number of samples on various hardware}
\label{tab:experiment_summary}
\footnotesize
\sisetup{round-mode=places,separate-uncertainty,
detect-weight=true,detect-inline-weight=math
}
\begin{tabular}{cccc
S[round-mode=uncertainty,round-precision=2,round-pad=false]
cccc
S[round-mode=uncertainty,round-precision=2,round-pad=false]
}
\toprule
\text{CPU} & \text{GPU} & \text{Samples} & \text{Method} & \text{Avg. Time [\si{\ms}]} & \text{CPU} & \text{GPU} & \text{Samples} & \text{Method} & \text{Avg. Time [\si{\ms}]} \\ \cmidrule(r){1-5} \cmidrule(l){6-10}
Intel 13900K & 1060 6GB & 128 & MPPI-Generic & \num{0.180181 \pm 0.0347118} & Intel 13900K & 1080 Ti & 128 & MPPI-Generic & \num{0.179371 \pm 0.0289556} \\
Intel 13900K & 1060 6GB & 128 & autorally & \num{0.436403 \pm 0.0541172}  & Intel 13900K & 1080 Ti & 128 & autorally & \num{0.443172 \pm 0.115946} \\
Intel 13900K & 1060 6GB & 128 & torchrl & \num{24.55619597 \pm 1.644165462} & Intel 13900K & 1080 Ti & 128 & torchrl & \num{24.62926912 \pm 1.805008154} \\
\cmidrule(r){1-5} \cmidrule(l){6-10}
Intel 13900K & 1060 6GB & 256 & MPPI-Generic & \num{0.191272 \pm 0.031385} & Intel 13900K & 1080 Ti & 256 & MPPI-Generic & \num{0.184424 \pm 0.0294887} \\
Intel 13900K & 1060 6GB & 256 & autorally & \num{0.431935 \pm 0.0372634} & Intel 13900K & 1080 Ti & 256 & autorally & \num{0.438665 \pm 0.0359797} \\
Intel 13900K & 1060 6GB & 256 & torchrl & \num{28.41556668 \pm 1.401850333} & Intel 13900K & 1080 Ti & 256 & torchrl & \num{28.40440059 \pm 1.401767178} \\
\cmidrule(r){1-5} \cmidrule(l){6-10}
Intel 13900K & 1060 6GB & 512 & MPPI-Generic & \num{0.225903 \pm 0.0314702} & Intel 13900K & 1080 Ti & 512 & MPPI-Generic & \num{0.200678 \pm 0.0306286} \\
Intel 13900K & 1060 6GB & 512 & autorally & \num{0.434337 \pm 0.0393827} & Intel 13900K & 1080 Ti & 512 & autorally & \num{0.442098 \pm 0.0376132} \\
Intel 13900K & 1060 6GB & 512 & torchrl & \num{28.56920576 \pm 1.625594162} & Intel 13900K & 1080 Ti & 512 & torchrl & \num{28.44526076 \pm 1.600661798} \\
\cmidrule(r){1-5} \cmidrule(l){6-10}
Intel 13900K & 1060 6GB & 1024 & MPPI-Generic & \num{0.310363 \pm 0.0345482} & Intel 13900K & 1080 Ti & 1024 & MPPI-Generic & \num{0.219663 \pm 0.0297783} \\
Intel 13900K & 1060 6GB & 1024 & autorally & \num{0.467177 \pm 0.0427389} & Intel 13900K & 1080 Ti & 1024 & autorally & \num{0.447832 \pm 0.0379423} \\
Intel 13900K & 1060 6GB & 1024 & torchrl & \num{28.36999822 \pm 1.712465173} & Intel 13900K & 1080 Ti & 1024 & torchrl & \num{28.53598952 \pm 1.599407986} \\
\cmidrule(r){1-5} \cmidrule(l){6-10}
Intel 13900K & 1060 6GB & 2048 & MPPI-Generic & \num{0.479197 \pm 0.0393547} & Intel 13900K & 1080 Ti & 2048 & MPPI-Generic & \num{0.297774 \pm 0.0333149} \\
Intel 13900K & 1060 6GB & 2048 & autorally & \num{0.544643 \pm 0.0671926} & Intel 13900K & 1080 Ti & 2048 & autorally & \num{0.481744 \pm 0.0412484} \\
Intel 13900K & 1060 6GB & 2048 & torchrl & \num{28.59234643 \pm 1.469444317} & Intel 13900K & 1080 Ti & 2048 & torchrl & \num{28.65779209 \pm 1.390036982} \\
\cmidrule(r){1-5} \cmidrule(l){6-10}
Intel 13900K & 1060 6GB & 4096 & MPPI-Generic & \num{0.890357 \pm 0.0535622} & Intel 13900K & 1080 Ti & 4096 & MPPI-Generic & \num{0.432359 \pm 0.0404817} \\
Intel 13900K & 1060 6GB & 4096 & autorally & \num{0.986041 \pm 0.0777549} & Intel 13900K & 1080 Ti & 4096 & autorally & \num{0.548638 \pm 0.0420299} \\
Intel 13900K & 1060 6GB & 4096 & torchrl & \num{29.32528877 \pm 1.707160026} & Intel 13900K & 1080 Ti & 4096 & torchrl & \num{28.90866542 \pm 1.690516162} \\
\cmidrule(r){1-5} \cmidrule(l){6-10}
Intel 13900K & 1060 6GB & 6144 & autorally & \num{1.1427 \pm 0.0774047} & Intel 13900K & 1080 Ti & 6144 & autorally & \num{0.61948 \pm 0.0409167} \\
Intel 13900K & 1060 6GB & 6144 & MPPI-Generic & \num{1.31063 \pm 0.0632859} & Intel 13900K & 1080 Ti & 6144 & MPPI-Generic & \num{0.701594 \pm 0.0459631} \\
Intel 13900K & 1060 6GB & 6144 & torchrl & \num{30.90311837 \pm 1.655715708}& Intel 13900K & 1080 Ti & 6144 & torchrl & \num{29.17314219 \pm 1.68587023} \\
\cmidrule(r){1-5} \cmidrule(l){6-10}
Intel 13900K & 1060 6GB & 8192 & autorally & \num{1.59904 \pm 0.0712165} & Intel 13900K & 1080 Ti & 8192 & autorally & \num{0.695471 \pm 0.0472906} \\
Intel 13900K & 1060 6GB & 8192 & MPPI-Generic & \num{1.75541 \pm 0.0702034} & Intel 13900K & 1080 Ti & 8192 & MPPI-Generic & \num{1.10935 \pm 0.0576267} \\
Intel 13900K & 1060 6GB & 8192 & torchrl & \num{31.71687722 \pm 1.393955124} & Intel 13900K & 1080 Ti & 8192 & torchrl & \num{29.74307036 \pm 1.505358741} \\
\cmidrule(r){1-5} \cmidrule(l){6-10}
Intel 13900K & 1060 6GB & 16384 & autorally & \num{3.04568 \pm 0.635748} & Intel 13900K & 1080 Ti & 16384 & autorally & \num{1.33403 \pm 0.0605589} \\
Intel 13900K & 1060 6GB & 16384 & MPPI-Generic & \num{3.33723 \pm 0.098572} & Intel 13900K & 1080 Ti & 16384 & MPPI-Generic & \num{2.18785 \pm 0.0784581} \\
Intel 13900K & 1060 6GB & 16384 & torchrl & \num{41.26050401 \pm 1.609524793} & Intel 13900K & 1080 Ti & 16384 & torchrl & \num{33.75626612 \pm 1.568212997} \\
\midrule
\midrule
Intel 13900K & 2080 & 128 & MPPI-Generic & \num{0.165901 \pm 0.0231691} & Intel 13900K & 4090 & 128 & MPPI-Generic & \num{0.122728 \pm 0.0317635} \\
Intel 13900K & 2080 & 128 & autorally & \num{0.293212 \pm 0.137769} & Intel 13900K & 4090 & 128 & autorally & \num{0.197862 \pm 0.0389471} \\
Intel 13900K & 2080 & 128 & torchrl & \num{24.81715465 \pm 1.604587402} & Intel 13900K & 4090 & 128 & torchrl & \num{24.47159696 \pm 1.629851147} \\
\cmidrule(r){1-5} \cmidrule(l){6-10}
Intel 13900K & 2080 & 256 & MPPI-Generic & \num{0.171244 \pm 0.0240126} &Intel 13900K & 4090 & 256 & MPPI-Generic & \num{0.121906 \pm 0.0287888} \\
Intel 13900K & 2080 & 256 & autorally & \num{0.291783 \pm 0.0281587} & Intel 13900K & 4090 & 256 & autorally & \num{0.196282 \pm 0.0291502} \\
Intel 13900K & 2080 & 256 & torchrl & \num{28.63221335 \pm 1.294271899} & Intel 13900K & 4090 & 256 & torchrl & \num{28.16628242 \pm 1.328607151} \\
\cmidrule(r){1-5} \cmidrule(l){6-10}
Intel 13900K & 2080 & 512 & MPPI-Generic & \num{0.180227 \pm 0.0280305} & Intel 13900K & 4090 & 512 & MPPI-Generic & \num{0.123621 \pm 0.0264929} \\
Intel 13900K & 2080 & 512 & autorally & \num{0.299438 \pm 0.0315328} & Intel 13900K & 4090 & 512 & autorally & \num{0.197492 \pm 0.0280592} \\
Intel 13900K & 2080 & 512 & torchrl & \num{28.70388794 \pm 1.54553658} & Intel 13900K & 4090 & 512 & torchrl & \num{28.22097349 \pm 1.604033825} \\
\cmidrule(r){1-5} \cmidrule(l){6-10}
Intel 13900K & 2080 & 1024 & MPPI-Generic & \num{0.203835 \pm 0.0278323} & Intel 13900K & 4090 & 1024 & MPPI-Generic & \num{0.13065 \pm 0.0252133} \\
Intel 13900K & 2080 & 1024 & autorally & \num{0.302238 \pm 0.028387} & Intel 13900K & 4090 & 1024 & autorally & \num{0.202446 \pm 0.0246008} \\
Intel 13900K & 2080 & 1024 & torchrl & \num{28.76341462 \pm 1.535270445} & Intel 13900K & 4090 & 1024 & torchrl & \num{28.13660431 \pm 1.556810549} \\
\cmidrule(r){1-5} \cmidrule(l){6-10}
Intel 13900K & 2080 & 2048 & MPPI-Generic & \num{0.260856 \pm 0.02615} & Intel 13900K & 4090 & 2048 & MPPI-Generic & \num{0.146179 \pm 0.0362073} \\
Intel 13900K & 2080 & 2048 & autorally & \num{0.32263 \pm 0.0333094} & Intel 13900K & 4090 & 2048 & autorally & \num{0.213538 \pm 0.0345784} \\
Intel 13900K & 2080 & 2048 & torchrl & \num{28.82683158 \pm 1.342341806} & Intel 13900K & 4090 & 2048 & torchrl & \num{28.20549726 \pm 1.322743723} \\
\cmidrule(r){1-5} \cmidrule(l){6-10}
Intel 13900K & 2080 & 4096 & MPPI-Generic & \num{0.397538 \pm 0.0344002} & Intel 13900K & 4090 & 4096 & MPPI-Generic & \num{0.169998 \pm 0.0295232} \\
Intel 13900K & 2080 & 4096 & autorally & \num{0.40302 \pm 0.0347713} & Intel 13900K & 4090 & 4096 & autorally & \num{0.22406 \pm 0.0283723} \\
Intel 13900K & 2080 & 4096 & torchrl & \num{29.09921718 \pm 1.59030902} & Intel 13900K & 4090 & 4096 & torchrl & \num{28.24524951 \pm 1.561157368} \\
\cmidrule(r){1-5} \cmidrule(l){6-10}
Intel 13900K & 2080 & 6144 & autorally & \num{0.518005 \pm 0.0398341} & Intel 13900K & 4090 & 6144 & MPPI-Generic & \num{0.200644 \pm 0.0276761} \\
Intel 13900K & 2080 & 6144 & MPPI-Generic & \num{0.576165 \pm 0.0423996} & Intel 13900K & 4090 & 6144 & autorally & \num{0.232756 \pm 0.0321019} \\
Intel 13900K & 2080 & 6144 & torchrl & \num{29.13836265 \pm 1.585337189} & Intel 13900K & 4090 & 6144 & torchrl & \num{28.35752892 \pm 1.587800846} \\
\cmidrule(r){1-5} \cmidrule(l){6-10}
Intel 13900K & 2080 & 8192 & autorally & \num{0.580367 \pm 0.0436748} & Intel 13900K & 4090 & 8192 & MPPI-Generic & \num{0.231558 \pm 0.0295746} \\
Intel 13900K & 2080 & 8192 & MPPI-Generic & \num{0.704521 \pm 0.0427607} & Intel 13900K & 4090 & 8192 & autorally & \num{0.245015 \pm 0.0265846} \\
Intel 13900K & 2080 & 8192 & torchrl & \num{29.41722679 \pm 1.276422812} & Intel 13900K & 4090 & 8192 & torchrl & \num{28.59068131 \pm 1.50733815} \\
\cmidrule(r){1-5} \cmidrule(l){6-10}
Intel 13900K & 2080 & 16384 & autorally & \num{1.19245 \pm 0.0593496} & Intel 13900K & 4090 & 16384 & autorally & \num{0.338403 \pm 0.0326836} \\
Intel 13900K & 2080 & 16384 & MPPI-Generic & \num{1.41827 \pm 0.0613242} & Intel 13900K & 4090 & 16384 & MPPI-Generic & \num{0.388618 \pm 0.038214} \\
Intel 13900K & 2080 & 16384 & torchrl & \num{32.36972547 \pm 1.529793824} & Intel 13900K & 4090 & 16384 & torchrl & \num{28.60460329 \pm 1.627474615} \\
\bottomrule
\end{tabular}
\end{table*}

\begin{table*}[hb!]
\centering

\label{tab:experiment_summary_part2}
\footnotesize
\sisetup{round-mode=places,separate-uncertainty,
detect-weight=true,detect-inline-weight=math
}
\begin{tabular}{cccc
S[round-mode=uncertainty,round-precision=3,round-pad=false]
cccc
S[round-mode=uncertainty,round-precision=3,round-pad=false]
}
\toprule
\text{CPU} & \text{GPU} & \text{Samples} & \text{Method} & \text{Avg. Time [\si{\ms}]} & \text{CPU} & \text{GPU} & \text{Samples} & \text{Method} & \text{Avg. Time [\si{\ms}]} \\ \cmidrule(r){1-5} \cmidrule(l){6-10}
AMD 5600X & 3080 & 128 & MPPI-Generic & \num{0.155392 \pm 0.0155994} & Intel 13900K & 3080 & 128 & MPPI-Generic & \num{0.147546 \pm 0.0316142} \\
AMD 5600X & 3080 & 128 & autorally & \num{0.270118 \pm 0.0169162} & Intel 13900K & 3080 & 128 & autorally & \num{0.245459 \pm 0.118456} \\
AMD 5600X & ~ & 128 & nav2 & \num{0.615263 \pm 0.0319062} & Intel 13900K & ~ & 128 & nav2 & \num{0.389923 \pm 0.0162476} \\
AMD 5600X & 3080 & 128 & torchrl & \num{47.61079502 \pm 1.583180632} & Intel 13900K & 3080 & 128 & torchrl & \num{24.77447772 \pm 1.728524408} \\
\cmidrule(r){1-5} \cmidrule(l){6-10}
AMD 5600X & 3080 & 256 & MPPI-Generic & \num{0.153569 \pm 0.0335784} &  Intel 13900K & 3080 & 256 & MPPI-Generic & \num{0.152581 \pm 0.027063} \\
AMD 5600X & 3080 & 256 & autorally & \num{0.274294 \pm 0.0151565} &  Intel 13900K & 3080 & 256 & autorally & \num{0.242785 \pm 0.029394} \\
AMD 5600X & ~ & 256 & nav2 & \num{1.08037 \pm 0.0426558} & Intel 13900K & ~ & 256 & nav2 & \num{0.728985 \pm 0.0101121} \\
AMD 5600X & 3080 & 256 & torchrl & \num{53.87716532 \pm 2.02432958} & Intel 13900K & 3080 & 256 & torchrl & \num{28.48625875 \pm 1.450323227} \\
\cmidrule(r){1-5} \cmidrule(l){6-10}
AMD 5600X & 3080 & 512 & MPPI-Generic & \num{0.152786 \pm 0.0115133} & Intel 13900K & 3080 & 512 & MPPI-Generic & \num{0.153557 \pm 0.0344778} \\
AMD 5600X & 3080 & 512 & autorally & \num{0.26958 \pm 0.0123445} & Intel 13900K & 3080 & 512 & autorally & \num{0.246625 \pm 0.0368777} \\
AMD 5600X & ~ & 512 & nav2 & \num{2.10114 \pm 0.0961835} & Intel 13900K & ~ & 512 & nav2 & \num{1.41862 \pm 0.0445258} \\
AMD 5600X & 3080 & 512 & torchrl & \num{53.88968277 \pm 2.021575269} & Intel 13900K & 3080 & 512 & torchrl & \num{28.58349538 \pm 1.653015595} \\
\cmidrule(r){1-5} \cmidrule(l){6-10}
AMD 5600X & 3080 & 1024 & MPPI-Generic & \num{0.170408 \pm 0.0154526} & Intel 13900K & 3080 & 1024 & MPPI-Generic & \num{0.16343 \pm 0.0271352} \\
AMD 5600X & 3080 & 1024 & autorally & \num{0.275659 \pm 0.0147643} & Intel 13900K & 3080 & 1024 & autorally & \num{0.259898 \pm 0.0357351} \\
AMD 5600X & ~ & 1024 & nav2 & \num{4.75178 \pm 0.240294} & Intel 13900K & ~ & 1024 & nav2 & \num{3.35586 \pm 0.0350171} \\
AMD 5600X & 3080 & 1024 & torchrl & \num{53.96552968 \pm 1.80221912} & Intel 13900K & 3080 & 1024 & torchrl & \num{28.67716479 \pm 1.746510696} \\
\cmidrule(r){1-5} \cmidrule(l){6-10}
AMD 5600X & 3080 & 2048 & MPPI-Generic & \num{0.197253 \pm 0.013828} & Intel 13900K & 3080 & 2048 & MPPI-Generic & \num{0.191848 \pm 0.03305} \\
AMD 5600X & 3080 & 2048 & autorally & \num{0.285042 \pm 0.0142588} & Intel 13900K & 3080 & 2048 & autorally & \num{0.262051 \pm 0.0389669} \\
AMD 5600X & ~ & 2048 & nav2 & \num{9.49957 \pm 0.379606} & Intel 13900K & ~ & 2048 & nav2 & \num{6.72117 \pm 0.0650638}  \\
AMD 5600X & 3080 & 2048 & torchrl & \num{54.02042031 \pm 2.00166666} & Intel 13900K & 3080 & 2048 & torchrl & \num{28.67557049 \pm 1.463939992} \\
\cmidrule(r){1-5} \cmidrule(l){6-10}
AMD 5600X & 3080 & 4096 & MPPI-Generic & \num{0.256218 \pm 0.0376908} & Intel 13900K & 3080 & 4096 & MPPI-Generic & \num{0.248872 \pm 0.0388457} \\
AMD 5600X & 3080 & 4096 & autorally & \num{0.289591 \pm 0.0142701} & Intel 13900K & 3080 & 4096 & autorally & \num{0.275967 \pm 0.0425595} \\
AMD 5600X & ~ & 4096 & nav2 & \num{19.5829 \pm 0.78882} & Intel 13900K & ~ & 4096 & nav2 & \num{13.9169 \pm 0.151997}  \\
AMD 5600X & 3080 & 4096 & torchrl & \num{54.14841008 \pm 2.084040739} & Intel 13900K & 3080 & 4096 & torchrl & \num{28.75650024 \pm 1.631511033} \\
\cmidrule(r){1-5} \cmidrule(l){6-10}
AMD 5600X & 3080 & 6144 & MPPI-Generic & \num{0.314502 \pm 0.0194076} & Intel 13900K & 3080 & 6144 & MPPI-Generic & \num{0.306363 \pm 0.0391016} \\
AMD 5600X & 3080 & 6144 & autorally & \num{0.336179 \pm 0.0142641} & Intel 13900K & 3080 & 6144 & autorally & \num{0.324317 \pm 0.0430433} \\
AMD 5600X & ~ & 6144 & nav2 & \num{29.8719 \pm 0.864029} & Intel 13900K & ~ & 6144 & nav2 & \num{21.7392 \pm 0.188897}  \\
AMD 5600X & 3080 & 6144 & torchrl & \num{54.18311858 \pm 1.787616743} & Intel 13900K & 3080 & 6144 & torchrl & \num{28.83032084 \pm 1.647170024} \\
\cmidrule(r){1-5} \cmidrule(l){6-10}
AMD 5600X & 3080 & 8192 & autorally & \num{0.366365 \pm 0.0167136} & Intel 13900K & 3080 & 8192 & MPPI-Generic & \num{0.35896 \pm 0.0371319} \\
AMD 5600X & 3080 & 8192 & MPPI-Generic & \num{0.369283 \pm 0.0210465} & Intel 13900K & 3080 & 8192 & autorally & \num{0.360461 \pm 0.0448512} \\
AMD 5600X & ~ & 8192 & nav2 & \num{40.6365 \pm 1.05923} & Intel 13900K & 3080 & 8192 & torchrl & \num{28.978863 \pm 1.443538958} \\
AMD 5600X & 3080 & 8192 & torchrl & \num{54.19890475 \pm 2.093610932} & Intel 13900K & ~ & 8192 & nav2 & \num{29.9169 \pm 0.239302}  \\
\cmidrule(r){1-5} \cmidrule(l){6-10}
AMD 5600X & 3080 & 16384 & autorally & \num{0.512088 \pm 0.0182601} & Intel 13900K & 3080 & 16384 & autorally & \num{0.499134 \pm 0.0497034} \\
AMD 5600X & 3080 & 16384 & MPPI-Generic & \num{0.578981 \pm 0.0240727} & Intel 13900K & 3080 & 16384 & MPPI-Generic & \num{0.627083 \pm 0.0474846} \\
AMD 5600X & 3080 & 16384 & torchrl & \num{54.67918301 \pm 1.695322846} & Intel 13900K & 3080 & 16384 & torchrl & \num{29.97758317 \pm 1.674777268} \\
AMD 5600X & ~ & 16384 & nav2 & \num{84.2601 \pm 1.64613} & Intel 13900K & ~ & 16384 & nav2 & \num{62.2642 \pm 0.990145}  \\
\midrule
\midrule
Jetson Nano & Tegra X1 & 128 & autorally & \num{1.39656 \pm 0.613235} & Jetson Nano & Tegra X1 & 128 & MPPI-Generic & \num{1.02439 \pm 0.0914058} \\
Jetson Nano & Tegra X1 & 256 & autorally & \num{1.50645 \pm 0.575889} & Jetson Nano & Tegra X1 & 256 & MPPI-Generic & \num{1.38392 \pm 0.0613626} \\
Jetson Nano & Tegra X1 & 512 & autorally & \num{1.84651 \pm 0.574576} & Jetson Nano & Tegra X1 & 512 & MPPI-Generic & \num{2.12254 \pm 0.11991} \\
Jetson Nano & Tegra X1 & 1024 & autorally & \num{3.24459 \pm 0.607794} & Jetson Nano & Tegra X1 & 1024 & MPPI-Generic & \num{4.10541 \pm 1.20212} \\
Jetson Nano & Tegra X1 & 2048 & autorally & \num{5.921 \pm 0.577712} & Jetson Nano & Tegra X1 & 2048 & MPPI-Generic & \num{7.29689 \pm 0.398892} \\
Jetson Nano & Tegra X1 & 4096 & autorally & \num{11.1589 \pm 0.581031} & Jetson Nano & Tegra X1 & 4096 & MPPI-Generic & \num{14.1242 \pm 0.133369} \\
Jetson Nano & Tegra X1 & 6144 & autorally & \num{16.5458 \pm 0.605249} & Jetson Nano & Tegra X1 & 6144 & MPPI-Generic & \num{21.0746 \pm 0.229213} \\
Jetson Nano & Tegra X1 & 8192 & autorally & \num{21.7672 \pm 0.628532} & Jetson Nano & Tegra X1 & 8192 & MPPI-Generic & \num{28.0223 \pm 0.260565} \\
Jetson Nano & Tegra X1 & 16384 & autorally & \num{44.1579 \pm 0.947064} & Jetson Nano & Tegra X1 & 16384 & MPPI-Generic & \num{56.6497 \pm 0.350869} \\
\midrule
\midrule
\multicolumn{2}{c}{Jetson Orin Nano (8 GB)} & 128 & MPPI-Generic & \num{0.629243 \pm 0.1296} & \multicolumn{2}{c}{Jetson Orin Nano (8 GB)} & 4096 & autorally & \num{2.71273 \pm 0.124459}\\
\multicolumn{2}{c}{Jetson Orin Nano (8 GB)} & 128 & autorally & \num{0.946763 \pm 0.292363} & \multicolumn{2}{c}{Jetson Orin Nano (8 GB)} & 4096 & MPPI-Generic & \num{3.09244 \pm 0.0791065} \\
\multicolumn{2}{c}{Jetson Orin Nano (8 GB)} & 128 & nav2 & \num{2.70737 \pm 0.0213212} & \multicolumn{2}{c}{Jetson Orin Nano (8 GB)} & 4096 & nav2 & \num{87.3834 \pm 0.260187} \\
\multicolumn{2}{c}{Jetson Orin Nano (8 GB)} & 128 & torchrl & \num{244.52353 \pm 6.83638} & \multicolumn{2}{c}{Jetson Orin Nano (8 GB)} & 4096 & torchrl & \num{278.75793 \pm 6.93590} \\
\cmidrule(r){1-5} \cmidrule(l){6-10}
\multicolumn{2}{c}{Jetson Orin Nano (8 GB)} & 256 & MPPI-Generic & \num{0.607396 \pm 0.0221034} & \multicolumn{2}{c}{Jetson Orin Nano (8 GB)} & 6144 & autorally & \num{3.86298 \pm 0.100405} \\
\multicolumn{2}{c}{Jetson Orin Nano (8 GB)} & 256 & autorally & \num{0.90375 \pm 0.0666886} & \multicolumn{2}{c}{Jetson Orin Nano (8 GB)} & 6144 & MPPI-Generic & \num{4.54514 \pm 0.135722} \\
\multicolumn{2}{c}{Jetson Orin Nano (8 GB)} & 256 & nav2 & \num{5.14864 \pm 0.0638036} & \multicolumn{2}{c}{Jetson Orin Nano (8 GB)} & 6144 & nav2 & \num{129.927 \pm 0.347753} \\
\multicolumn{2}{c}{Jetson Orin Nano (8 GB)} & 256 & torchrl & \num{272.14235 \pm 8.28430} & \multicolumn{2}{c}{Jetson Orin Nano (8 GB)} & 6144 & torchrl & \num{283.59609 \pm 8.08859} \\
\cmidrule(r){1-5} \cmidrule(l){6-10}
\multicolumn{2}{c}{Jetson Orin Nano (8 GB)} & 512 & MPPI-Generic & \num{0.728221 \pm 0.0301545} & \multicolumn{2}{c}{Jetson Orin Nano (8 GB)} & 8192 & autorally & \num{5.25501 \pm 0.131005} \\
\multicolumn{2}{c}{Jetson Orin Nano (8 GB)} & 512 & autorally & \num{0.909514 \pm 0.0332163} & \multicolumn{2}{c}{Jetson Orin Nano (8 GB)} & 8192 & MPPI-Generic & \num{6.10201 \pm 0.153504} \\
\multicolumn{2}{c}{Jetson Orin Nano (8 GB)} & 512 & nav2 & \num{10.0513 \pm 0.0571951} & \multicolumn{2}{c}{Jetson Orin Nano (8 GB)} & 8192 & nav2 & \num{170.249 \pm 0.356299} \\
\multicolumn{2}{c}{Jetson Orin Nano (8 GB)} & 512 & torchrl & \num{272.298624 \pm 6.71430} & \multicolumn{2}{c}{Jetson Orin Nano (8 GB)} & 8192 & torchrl & \num{289.85103 \pm 6.66031} \\
\cmidrule(r){1-5} \cmidrule(l){6-10}
\multicolumn{2}{c}{Jetson Orin Nano (8 GB)} & 1024 & MPPI-Generic & \num{1.03002 \pm 0.0349694} & \multicolumn{2}{c}{Jetson Orin Nano (8 GB)} & 16384 & autorally & \num{9.94003 \pm 0.196983} \\
\multicolumn{2}{c}{Jetson Orin Nano (8 GB)} & 1024 & autorally & \num{1.03162 \pm 0.0307838} & \multicolumn{2}{c}{Jetson Orin Nano (8 GB)} & 16384 & MPPI-Generic & \num{12.0695 \pm 0.217526} \\
\multicolumn{2}{c}{Jetson Orin Nano (8 GB)} & 1024 & nav2 & \num{21.4881 \pm 0.10538} & \multicolumn{2}{c}{Jetson Orin Nano (8 GB)} & 16384 & torchrl & \num{314.67959 \pm 8.52167} \\
\multicolumn{2}{c}{Jetson Orin Nano (8 GB)} & 1024 & torchrl & \num{273.90642 \pm 8.26401} & \multicolumn{2}{c}{Jetson Orin Nano (8 GB)} & 16384 & nav2 & \num{334.911 \pm 0.74598} \\
\cmidrule(r){1-5} \cmidrule(l){6-10}
\multicolumn{2}{c}{Jetson Orin Nano (8 GB)} & 2048 & autorally & \num{1.40894 \pm 0.0556146} & & & & \\
\multicolumn{2}{c}{Jetson Orin Nano (8 GB)} & 2048 & MPPI-Generic & \num{1.64541 \pm 0.108883} & & & & \\
\multicolumn{2}{c}{Jetson Orin Nano (8 GB)} & 2048 & nav2 & \num{43.7964 \pm 0.186287} & & & & \\
\multicolumn{2}{c}{Jetson Orin Nano (8 GB)} & 2048 & torchrl & \num{275.45777 \pm 8.23631} & & & & \\
\bottomrule
\end{tabular}
\end{table*}

\begin{table*}[hb!]
\centering

\label{tab:experiment_summary_part3}
\footnotesize
\sisetup{round-mode=places,separate-uncertainty,
detect-weight=true,detect-inline-weight=math
}
\begin{tabular}{cccc
S[round-mode=uncertainty,round-precision=3,round-pad=false,tight-spacing=true]
cccc
S[round-mode=uncertainty,round-precision=3,round-pad=false]
}
\toprule
\text{CPU} & \text{GPU} & \text{Samples} & \text{Method} & \text{Avg. Time [\si{\ms}]} & \text{CPU} & \text{GPU} & \text{Samples} & \text{Method} & \text{Avg. Time [\si{\ms}]} \\ \cmidrule(r){1-5} \cmidrule(l){6-10}

AMD 5600X & 1050 Ti & MPPI-Generic & 128 & \num{0.214853 \pm 0.0478001} & Intel 13900K & 1050 Ti & MPPI-Generic & 128 & \num{0.200793 \pm 0.059054} \\
AMD 5600X & 1050 Ti & autorally & 128 & \num{0.464578 \pm 0.0336889} & Intel 13900K & 1050 Ti & autorally & 128 & \num{0.455909 \pm 0.159941} \\
AMD 5600X & 1050 Ti & torchrl & 128 & \num{46.64942789077753 \pm 2.249891630328609} & Intel 13900K & 1050 Ti & torchrl & 128 & \num{25.603788137435888 \pm 2.7138171474565613} \\
\cmidrule(r){1-5} \cmidrule(l){6-10}
AMD 5600X & 1050 Ti & MPPI-Generic & 256 & \num{0.226453 \pm 0.0433133} & Intel 13900K & 1050 Ti & MPPI-Generic & 256 & \num{0.225711 \pm 0.0578007} \\
AMD 5600X & 1050 Ti & autorally & 256 & \num{0.464905 \pm 0.0338034} & Intel 13900K & 1050 Ti & autorally & 256 & \num{0.454321 \pm 0.0832647} \\
AMD 5600X & 1050 Ti & torchrl & 256 & \num{52.884384155273416 \pm 2.402423955959606} & Intel 13900K & 1050 Ti & torchrl & 256 & \num{29.307887554168733 \pm 2.4366536593302537} \\
\cmidrule(r){1-5} \cmidrule(l){6-10}
AMD 5600X & 1050 Ti & MPPI-Generic & 512 & \num{0.307676 \pm 0.052547} & Intel 13900K & 1050 Ti & MPPI-Generic & 512 & \num{0.301609 \pm 0.0657743} \\
AMD 5600X & 1050 Ti & autorally & 512 & \num{0.481166 \pm 0.0374961} & Intel 13900K & 1050 Ti & autorally & 512 & \num{0.468811 \pm 0.0801855} \\
AMD 5600X & 1050 Ti & torchrl & 512 & \num{52.96501350402835 \pm 2.365726109028937} & Intel 13900K & 1050 Ti & torchrl & 512 & \num{29.90781569480895 \pm 2.7988644985867723} \\
\cmidrule(r){1-5} \cmidrule(l){6-10}
AMD 5600X & 1050 Ti & MPPI-Generic & 1024 & \num{0.437319 \pm 0.0548496} & Intel 13900K & 1050 Ti & MPPI-Generic & 1024 & \num{0.429377 \pm 0.0753275} \\
AMD 5600X & 1050 Ti & autorally & 1024 & \num{0.519615 \pm 0.0397184} & Intel 13900K & 1050 Ti & autorally & 1024 & \num{0.51198 \pm 0.0962865} \\
AMD 5600X & 1050 Ti & torchrl & 1024 & \num{53.14957308769222 \pm 2.144392550399996} & Intel 13900K & 1050 Ti & torchrl & 1024 & \num{29.90231370925901 \pm 2.9952270005719606} \\
\cmidrule(r){1-5} \cmidrule(l){6-10}
AMD 5600X & 1050 Ti & autorally & 2048 & \num{0.637062 \pm 0.0432679} & Intel 13900K & 1050 Ti & autorally & 2048 & \num{0.629705 \pm 0.104016} \\
AMD 5600X & 1050 Ti & MPPI-Generic & 2048 & \num{0.844269 \pm 0.0613633} & Intel 13900K & 1050 Ti & MPPI-Generic & 2048 & \num{0.83698 \pm 0.0993271} \\
AMD 5600X & 1050 Ti & torchrl & 2048 & \num{53.57521462440497 \pm 2.4539371433302883} & Intel 13900K & 1050 Ti & torchrl & 2048 & \num{30.38908171653747 \pm 2.918950472504109} \\
\cmidrule(r){1-5} \cmidrule(l){6-10}
AMD 5600X & 1050 Ti & autorally & 4096 & \num{1.29523 \pm 0.056928} & Intel 13900K & 1050 Ti & autorally & 4096 & \num{1.20165 \pm 0.132191} \\
AMD 5600X & 1050 Ti & MPPI-Generic & 4096 & \num{1.48358 \pm 0.0615682} & Intel 13900K & 1050 Ti & MPPI-Generic & 4096 & \num{1.46812 \pm 0.134319} \\
AMD 5600X & 1050 Ti & torchrl & 4096 & \num{55.103763341903694 \pm 2.424383291043795} & Intel 13900K & 1050 Ti & torchrl & 4096 & \num{31.28236222267151 \pm 2.7081101960851095} \\
\cmidrule(r){1-5} \cmidrule(l){6-10}
AMD 5600X & 1050 Ti & autorally & 6144 & \num{1.45951 \pm 0.0617682} & Intel 13900K & 1050 Ti & autorally & 6144 & \num{1.7757 \pm 0.150588} \\
AMD 5600X & 1050 Ti & MPPI-Generic & 6144 & \num{2.1196 \pm 0.0726572} & Intel 13900K & 1050 Ti & MPPI-Generic & 6144 & \num{2.10306 \pm 0.155404} \\
AMD 5600X & 1050 Ti & torchrl & 6144 & \num{57.42968893051147 \pm 2.1360445891937356} & Intel 13900K & 1050 Ti & torchrl & 6144 & \num{34.26116275787351 \pm 2.7545863723875437} \\
\cmidrule(r){1-5} \cmidrule(l){6-10}
AMD 5600X & 1050 Ti & autorally & 8192 & \num{2.07623 \pm 0.0738657} & Intel 13900K & 1050 Ti & autorally & 8192 & \num{2.36026 \pm 0.169856} \\
AMD 5600X & 1050 Ti & MPPI-Generic & 8192 & \num{2.90615 \pm 0.084103} & Intel 13900K & 1050 Ti & MPPI-Generic & 8192 & \num{2.89317 \pm 0.247637} \\
AMD 5600X & 1050 Ti & torchrl & 8192 & \num{60.502972364425666 \pm 2.4482911577226276} & Intel 13900K & 1050 Ti & torchrl & 8192 & \num{37.0794641971588 \pm 2.684802959376344} \\
\cmidrule(r){1-5} \cmidrule(l){6-10}
AMD 5600X & 1050 Ti & autorally & 16384 & \num{4.10761 \pm 0.112963} & Intel 13900K & 1050 Ti & autorally & 16384 & \num{4.61068 \pm 0.254627} \\
AMD 5600X & 1050 Ti & MPPI-Generic & 16384 & \num{5.66518 \pm 0.126393} & Intel 13900K & 1050 Ti & MPPI-Generic & 16384 & \num{5.63798 \pm 0.418242} \\
AMD 5600X & 1050 Ti & torchrl & 16384 & \num{73.40314197540287 \pm 1.9306471921792019} & Intel 13900K & 1050 Ti & torchrl & 16384 & \num{51.973490238189676 \pm 2.884466781077533} \\
\midrule
\midrule
AMD 5600X & 1650 & MPPI-Generic & 128 & \num{0.145933 \pm 0.0207775} & Intel 13900K & 1650 & MPPI-Generic & 128 & \num{0.135448 \pm 0.0263319} \\
AMD 5600X & 1650 & autorally & 128 & \num{0.255503 \pm 0.0176416} & Intel 13900K & 1650 & autorally & 128 & \num{0.244764 \pm 0.110752} \\
AMD 5600X & 1650 & torchrl & 128 & \num{46.523932695388815 \pm 1.9606853912492164} & Intel 13900K & 1650 & torchrl & 128 & \num{24.70457673072812 \pm 1.7695407930321057} \\
\cmidrule(r){1-5} \cmidrule(l){6-10}
AMD 5600X & 1650 & MPPI-Generic & 256 & \num{0.152956 \pm 0.0188354} & Intel 13900K & 1650 & MPPI-Generic & 256 & \num{0.152638 \pm 0.0260099} \\
AMD 5600X & 1650 & autorally & 256 & \num{0.256436 \pm 0.023154} & Intel 13900K & 1650 & autorally & 256 & \num{0.238484 \pm 0.0284148} \\
AMD 5600X & 1650 & torchrl & 256 & \num{52.68726444244381 \pm 2.313450177948185} & Intel 13900K & 1650 & torchrl & 256 & \num{28.543481349945065 \pm 1.5188542036891182} \\
\cmidrule(r){1-5} \cmidrule(l){6-10}
AMD 5600X & 1650 & MPPI-Generic & 512 & \num{0.201461 \pm 0.0218164} & Intel 13900K & 1650 & MPPI-Generic & 512 & \num{0.196435 \pm 0.0320444} \\
AMD 5600X & 1650 & autorally & 512 & \num{0.264917 \pm 0.0226419} & Intel 13900K & 1650 & autorally & 512 & \num{0.238714 \pm 0.0249006} \\
AMD 5600X & 1650 & torchrl & 512 & \num{52.196183681488016 \pm 2.124205141479847} & Intel 13900K & 1650 & torchrl & 512 & \num{28.674972057342522 \pm 1.6570401222206943} \\
\cmidrule(r){1-5} \cmidrule(l){6-10}
AMD 5600X & 1650 & MPPI-Generic & 1024 & \num{0.28475 \pm 0.0293078} & Intel 13900K & 1650 & MPPI-Generic & 1024 & \num{0.286556 \pm 0.0346445} \\
AMD 5600X & 1650 & autorally & 1024 & \num{0.325993 \pm 0.0224642} & Intel 13900K & 1650 & autorally & 1024 & \num{0.305997 \pm 0.0306022} \\
AMD 5600X & 1650 & torchrl & 1024 & \num{52.32941579818724 \pm 1.7772938567851295} & Intel 13900K & 1650 & torchrl & 1024 & \num{28.783922195434577 \pm 1.6545542624780571} \\
\cmidrule(r){1-5} \cmidrule(l){6-10}
AMD 5600X & 1650 & autorally & 2048 & \num{0.424721 \pm 0.0251058} & Intel 13900K & 1650 & autorally & 2048 & \num{0.404684 \pm 0.0381439} \\
AMD 5600X & 1650 & MPPI-Generic & 2048 & \num{0.504621 \pm 0.0294767} & Intel 13900K & 1650 & MPPI-Generic & 2048 & \num{0.483436 \pm 0.0431012} \\
AMD 5600X & 1650 & torchrl & 2048 & \num{52.57812213897705 \pm 2.109717144827896} & Intel 13900K & 1650 & torchrl & 2048 & \num{29.134824991226214 \pm 1.4589429078950413} \\
\cmidrule(r){1-5} \cmidrule(l){6-10}
AMD 5600X & 1650 & autorally & 4096 & \num{0.806484 \pm 0.044721} & Intel 13900K & 1650 & autorally & 4096 & \num{0.792676 \pm 0.0619396} \\
AMD 5600X & 1650 & MPPI-Generic & 4096 & \num{1.0046 \pm 0.0447515} & Intel 13900K & 1650 & MPPI-Generic & 4096 & \num{0.968006 \pm 0.0636123} \\
AMD 5600X & 1650 & torchrl & 4096 & \num{53.69180130958555 \pm 2.080985489081914} & Intel 13900K & 1650 & torchrl & 4096 & \num{30.612709522247307 \pm 1.6963275569290448} \\
\cmidrule(r){1-5} \cmidrule(l){6-10}
AMD 5600X & 1650 & autorally & 6144 & \num{0.984068 \pm 0.0400747} & Intel 13900K & 1650 & autorally & 6144 & \num{0.941817 \pm 0.0623022} \\
AMD 5600X & 1650 & MPPI-Generic & 6144 & \num{1.22994 \pm 0.0437208} & Intel 13900K & 1650 & MPPI-Generic & 6144 & \num{1.1777 \pm 0.0637399} \\
AMD 5600X & 1650 & torchrl & 6144 & \num{55.52646160125737 \pm 1.7959738229272482} & Intel 13900K & 1650 & torchrl & 6144 & \num{32.39474034309386 \pm 1.7242355788852255} \\
\cmidrule(r){1-5} \cmidrule(l){6-10}
AMD 5600X & 1650 & autorally & 8192 & \num{1.38696 \pm 0.0475587} & Intel 13900K & 1650 & autorally & 8192 & \num{1.33433 \pm 0.0721079} \\
AMD 5600X & 1650 & MPPI-Generic & 8192 & \num{1.77403 \pm 0.0514672} & Intel 13900K & 1650 & MPPI-Generic & 8192 & \num{1.67893 \pm 0.0757958} \\
AMD 5600X & 1650 & torchrl & 8192 & \num{57.84410929679877 \pm 2.053263522175228} & Intel 13900K & 1650 & torchrl & 8192 & \num{34.515250444412175 \pm 1.45809072678584} \\
\cmidrule(r){1-5} \cmidrule(l){6-10}
AMD 5600X & 1650 & autorally & 16384 & \num{2.70391 \pm 0.0684633} & Intel 13900K & 1650 & autorally & 16384 & \num{2.59031 \pm 0.0999424} \\
AMD 5600X & 1650 & MPPI-Generic & 16384 & \num{3.29337 \pm 0.110754} & Intel 13900K & 1650 & MPPI-Generic & 16384 & \num{3.12843 \pm 0.12766} \\
AMD 5600X & 1650 & torchrl & 16384 & \num{68.91886115074149 \pm 1.7342905821987826} & Intel 13900K & 1650 & torchrl & 16384 & \num{46.41932129859922 \pm 1.730473429964794} \\
\bottomrule
\end{tabular}
\end{table*}